\documentclass[journal]{IEEEtran}

\usepackage{amsmath,amssymb}
\usepackage{cases}
\usepackage{tabularx}
\usepackage{multicol}
\usepackage{graphicx}
\usepackage{float}
\usepackage{amsmath}
\usepackage{amssymb}
\usepackage{array}
\usepackage{color}
\usepackage{xcolor}
\usepackage[nocompress]{cite}
\usepackage{multirow}
\usepackage{color}
\usepackage{algorithmic}
\usepackage{bbm}
\usepackage{bm}
\usepackage{booktabs}
\newcommand{\specialcell}[2][c]{%
\begin{tabular}[#1]{@{}c@{}}#2\end{tabular}}
\graphicspath{{figures/}}
\newcolumntype{C}[1]{>{\centering\let\newline\\\arraybackslash\hspace{0pt}}m{#1}}

\newcommand{\cmt}[1]{ #1}
\usepackage[bookmarks,colorlinks,linkcolor=red, anchorcolor=blue, citecolor=green]{hyperref}
\usepackage{url}
\newcommand{\etal}{\textit{et al}.}
\newcommand{\ie}{\textit{i}.\textit{e}.}
\newcommand{\eg}{\textit{e}.\textit{g}.}

\newcommand{\CTX}{\mathrm{CTX}}

\newcommand{\GRP}{\mathrm{GP}}
\DeclareMathOperator*{\argmin}{argmin}
\graphicspath{{figs/}}

\usepackage{anyfontsize}
\newcommand{\mtiny}[1]{{\fontsize{4}{5}\selectfont #1}}

\begin{document}

\title{Efficient and Effective Context-Based Convolutional Entropy Modeling for Image Compression}

\author{Mu~Li,
        Kede~Ma,~\IEEEmembership{Member,~IEEE},
        Jane~You,~\IEEEmembership{Member,~IEEE},
        David~Zhang,~\IEEEmembership{Fellow,~IEEE},
        and~Wangmeng~Zuo,~\IEEEmembership{Senior Member,~IEEE}% <-this % stops a space
\IEEEcompsocitemizethanks{\IEEEcompsocthanksitem{This work is partially supported by the National Natural Scientific Foundation of China (NSFC) under Grant No. 61671182 and 61872118.}
\IEEEcompsocthanksitem Mu Li and Jane You are with the Department
of Computing, the Hong Kong Polytechnic University, Kowloon, Hong Kong (e-mail: csmuli@comp.polyu.edu.hk; csyjia@comp.polyu.edu.hk).
\IEEEcompsocthanksitem Wangmeng Zuo is with the School
of Computer Science and Technology, Harbin Institute of Technology, Harbin, 150001, China (e-mail: cswmzuo@gmail.com).
\IEEEcompsocthanksitem Kede Ma is with the Department
of Computer Science, City University of Hong Kong, Kowloon, Hong Kong (e-mail: kede.ma@cityu.edu.hk).
\IEEEcompsocthanksitem David Zhang is with the School of Science and Engineering, The Chinese University of Hong Kong (Shenzhen),  Shenzhen Research Institute of Big Data, and Shenzhen Institute of Artificial Intelligence and Robotics for Society, Shenzhen, China, e-mail: (davidzhang@cuhk.edu.cn).}% <-this % stops a space
% \thanks{Manuscript received xxx; revised xxx}
}

\markboth{}%
{Shell \MakeLowercase{\textit{et al.}}: Bare Demo of IEEEtran.cls for IEEE Journals}

\maketitle

\begin{abstract}
Precise estimation of the probabilistic structure of natural images plays an essential role in image compression.
Despite the recent remarkable success of end-to-end optimized image compression, the latent codes are usually assumed to be fully statistically factorized in order to simplify entropy modeling.
However, this assumption generally does not hold true and may hinder compression performance. Here we present context-based convolutional networks (CCNs) for efficient and effective entropy modeling.
In particular, a 3D zigzag scanning order and a 3D code dividing technique are introduced to define proper coding contexts for parallel entropy decoding, both of which boil down to place translation-invariant binary masks on convolution filters of CCNs.
We demonstrate the promise of CCNs for entropy modeling in both lossless and lossy image compression.
For the former, we directly apply a CCN to the binarized representation of an image to compute the Bernoulli distribution of each code for entropy estimation.
For the latter, the categorical distribution of each code is represented by a discretized mixture of Gaussian distributions, whose parameters are estimated by three CCNs. We then jointly optimize the CCN-based entropy model along with analysis and synthesis transforms for rate-distortion performance.
Experiments on the Kodak and Tecnick datasets show that our methods powered by the proposed CCNs generally achieve comparable
compression performance to the state-of-the-art while being much faster.
\end{abstract}


% \fi
\IEEEpeerreviewmaketitle

\section{Introduction}
Data compression has played a significant role in engineering for centuries~\cite{wiki:morse_code}. Compression can be either lossless or lossy. Lossless compression allows perfect data reconstruction from compressed bitstreams with the goal of assigning shorter codewords to more ``probable'' codes. Typical examples include Huffman coding~\cite{huffman1952method}, arithmetic coding~\cite{witten1987arithmetic}, and range coding~\cite{martin1979range}. Lossy compression discards ``unimportant'' information of the input data, and the definition of importance is application-dependent. For example, if the data (such as images and videos) are meant to be consumed by the human visual system, importance should be measured in accordance with human perception, discarding features that are perceptually redundant, while keeping those that are most visually noticeable.
In lossy compression, one must face the rate-distortion trade-off, where the rate is computed by the entropy of the discrete codes~\cite{Shannon1948} and the distortion is measured by a signal fidelity metric. A prevailing scheme in the context of lossy image compression is transform coding, which consists of three operations - transformation, quantization, and entropy coding. Transforms map an image to a latent code representation, which is better-suited for exploiting aspects of human perception. Early transforms~\cite{ahmed_1974_dct} are linear, invertible, and fixed for all bit rates; errors arise only from quantization. Recent transforms take the form of deep neural networks (DNNs)~\cite{balle2016end}, aiming for nonlinear and more compressible representations. DNN-based transforms are mostly non-invertible, which may, however, encourage discarding perceptually unimportant image features during transformation. This gives us an opportunity to learn different transforms at different bit rates for optimal rate-distortion performance. Entropy coding is responsible for losslessly compressing the quantized codes into bitstreams for storage and transmission.

In either lossless or lossy image compression, a discrete probability distribution of the latent codes shared by the encoder and the decoder (\ie, the entropy model) is essential in determining the compression performance. According to the Shannon's source coding theorem~\cite{Shannon1948}, given a vector of code intensities $\bm y =\{y_0,\ldots,y_M\}$, the optimal code length of $\bm y$ should be $\lceil-\log_{2}P(\bm y)\rceil$, where binary symbols are assumed to construct the codebook. Without further constraints, it is intractable to estimate $P(\bm y)$ in high-dimensional spaces, a problem commonly known as the curse of dimensionality. For this reason, most entropy coding schemes assume $\bm y$ is fully statistically factorized with the same marginal distribution, leading to a code length of $\lceil-\sum_{i=0}^{M}\log_{2}P(y_i)\rceil$. Alternatively, the chain rule in probability theory offers a more accurate approximation
\begin{align}
P(\bm y) \approx \prod_{i=0}^{M}P(y_i|\mathrm{PTX}(y_i,\bm y)),
\end{align}
where $\mathrm{PTX}(y_i,\bm y)\subset\lbrace y_0,\ldots, y_{i-1}\rbrace$ represents the partial context of $y_i$ coded before it in $\bm y$. A representative example is the context-based adaptive binary arithmetic coding (CABAC)~\cite{marpe2003context} in H.264/AVC, which considers the two nearest codes as partial context, and obtains noticeable improvements over previous image/video compression standards.
As the size of $\mathrm{PTX}(y_i,\bm y)$ becomes large, it is difficult to estimate this conditional probability by constructing histograms. Recent methods such as PixelRNN~\cite{oord2016pixel} and PixelCNN~\cite{oord2016conditional} take advantage of DNNs in modeling long range relations to increase the size of partial context, but are computationally intensive.

%, taking hours for medium-resolution images
%
%An acceleration version of Multiscale PixelCNN introduces some side information (a start-up image) and predict $2\times$ larger image from a start-up image and repeat the process to generate a large image.
%
%However, the introduced side information should be viewed as the codes too and will inevitably decrease the performance for entropy coding.

In this work, we present context-based convolutional networks (CCNs) for effective and efficient entropy modeling. Given $\bm y$, we specify a 3D zigzag coding order such that the most relevant codes of $y_i$ can be included in its context. Parallel computation during  entropy encoding is straightforward as the context of each code is known and readily available. However, this is not always the case during entropy decoding. The partial context of $y_i$ should first be decoded sequentially for the estimation of $P(y_i|\mathrm{PTX}(y_i,\bm y))$, which is prohibitively slow. To address this issue, we introduce a 3D code dividing technique, which partitions $\bm y$ into multiple groups in compliance with the proposed coding order. The codes within each group are assumed to be conditionally independent given their respective contexts, and therefore can be decoded in parallel. In the context of CCNs, this amounts to applying properly designed translation-invariant binary masks to convolutional filters.
%
%To improve the effectiveness of the entropy prediction with the partial context, a special designed coding order and code dividing scheme is proposed in this paper.
%
%

To validate the proposed CCNs, we combine them with arithmetic coding~\cite{witten1987arithmetic} for entropy modeling. For lossless image compression, we convert the input grayscale image to eight binary planes and train a CCN to predict the Bernoulli distribution of $y_i$ by optimizing the entropy loss in information theory~\cite{cover2006elements}.  For lossy image compression, we parameterize the categorical distribution of $y_i$ with a discretized mixture of Gaussian (MoG) distributions, whose parameters (\ie, mixture weights, means, and variances) are estimated by three CCNs. The CCN-based entropy model is jointly optimized with analysis and synthesis transforms (\ie, mappings between raw pixel space and latent code space) over a database of training images, trading off the rate and the distortion.
Experiments on the Kodak and Tecnick datasets show that our methods for lossless and lossy image compression perform favorably against image compression standards and DNN-based methods, especially at low bit rates.

% To conclude, the contributions of the paper are listed as follows:

% \begin{itemize}
% 	\item A set of slope convolutional networks have been proposed for context-based entropy modeling. With a special designed coding order and the code dividing scheme, the proposed slope convolutional network can perform parallel entropy prediction in both entropy encoding and decoding.
% 	\item A entropy coding scheme SCAE which combines the slope convolutional entropy prediction model with the entropy coding method arithmetic coding is introduced and can be used in any compression framework for context based entropy coding.  In this paper, the SCAE is directly trained to do entropy coding on binarized gray images and overwhelm all the current lossless image compression standards.
% 	\item A continuous context-based entropy loss is introduced for lossy image compression by supposing each code follows a MoG distribution with the parameters generated by the slope convolutional networks. With a joint ratio-distortion loss, the proposed lossy image compression model can be trained in an end-to-end manner and achieve state-of-art performance for lossy image compression application.
% \end{itemize}

\section{Related Work}
In this section, we provide a brief overview of entropy models and lossy image compression methods based on DNNs. For traditional image compression techniques, we refer interested readers to~\cite{sudhakar2005image,wallace1992jpeg,skodras2001jpeg}.

\subsection{DNN-Based Entropy Modeling}
 The first and the most important step in entropy modeling is to estimate the probability $P(\bm y)$. For most image compression techniques, $\bm y$ is assumed to be statistically independent, whose entropy can be easily computed through the marginal distributions~\cite{balle2016end,theis2017lossy,agustsson2017soft,rippel2017real}. Arguably speaking, natural images undergoing a highly nonlinear analysis transform still exhibit strong statistical redundancies~\cite{balle2018variational}. This suggests that incorporating context into probability estimation has great potentials in improving the performance of entropy coding.

 DNN-based context modeling for natural languages and images has attracted considerable attention in the past decade. In natural language processing, recurrent neural networks (RNN)~\cite{mikolov2010recurrent}, and long short-term memory (LSTM)~\cite{sundermeyer2012lstm} are two popular tools to model long-range dependencies. In image processing, PixelRNN~\cite{oord2016pixel} and PixelCNN~\cite{oord2016conditional} are among the first attempts to exploit long-range pixel dependencies for image generation.
The above-mentioned methods are computationally inefficient, requiring one forward propagation to generate (or estimate the probability of) a single pixel. To speed up PixelCNN, Reed~\etal~\cite{reed2017multi} proposed Multiscale PixelCNN, which is able to sample a twice larger intermediate image conditioning on the initial image. This process may be iterated to generate the final high-resolution result. When viewing Multiscale PixelCNN as an entropy model, we must losslessly compress and send the initial image as side information to the decoder for entropy decoding.

Only recently have DNNs for context-based entropy modeling become an active research topic.
Ball{\'e} \etal~\cite{balle2018variational} introduced a scale prior, which stores a variance parameter for each  $y_i$ as side information. Richer side information generally leads to more accurate entropy modeling. However, this type of  information should
also be quantized, compressed and considered as part of the codes, and it is difficult to trade off the bits saved by the improved entropy model and the bits introduced by storing this side information.
% Toderici \etal~\cite{toderici2016full} described a hybrid model comprised of RNNs, LSTMs, and CNNs for context-based entropy estimation.
Li \etal~\cite{li2017learning} extracted a small code block for each $y_i$  as its context, and adopted a simple DNN for entropy modeling. The method suffers from heavy computational complexity similar to PixelRNN~\cite{oord2016pixel}. Li \etal~\cite{li2018efficient} and Mentzer \etal~\cite{mentzer2018conditional1} implemented parallel entropy encoding with masked DNNs. However, sequential entropy decoding has to be performed due to the context dependence, which remains painfully slow.
%Mentzer et al. further used the masked FCN as the entropy loss to optimize the whole compression framework. Instead of adopting end-to-end training scheme, they alternatively optimize the entropy estimation FCN and the encoder-decoder network. As the entropy adopted in is discrete and nondifferentiable to the codes but differential to the context of the codes, they directly optimize the context of the codes instead of the codes themselves.
In contrast, our CCN-based entropy model permits parallel entropy encoding and decoding, making it more attractive for practical applications.
% And a continuous context-based entropy loss is introduced by supposing the codes follow of Mixture of Gaussian (MoG) distributions depending on the context where the parameter of the MoGs can be estimated from the codes with SCNs. With the continuous entropy, a encoder-decoder network can be trained with the joint ratio-distortion optimization.

\subsection{DNN-Based Lossy Image Compression}
A major problem in end-to-end lossy image compression is that the gradients of the  quantization function are zeros almost everywhere, making gradient descent-based optimization ineffective. Different strategies have been proposed to alleviate the zero-gradient problem resulting from quantization.
From a signal processing perspective, the quantizer can be approximated by additive i.i.d. uniform noise, which has the same width as the quantization bin~\cite{gray1998quantization}. A desired property of this approximation is that the resulting density is a continuous relaxation of the probability mass function of $\bm y$~\cite{balle2016end}.
Another line of research introduced continuous functions (without the zero-gradient problem) to approximate the quantization function.  The step quantizer is used in the forward pass, while its continuous proxy is used in the backward pass.
Toderici \etal~\cite{toderici2015variable} learned an  RNN  to compress small-size images  in a progressive manner. They later tested their models on large-size images~\cite{toderici2016full}. Johnston \etal~\cite{johnston2017improved} exploited adaptive bit allocations and  perceptual losses to boost the compression performance especially in terms of  MS-SSIM~\cite{wang2003multiscale}.

The joint optimization of rate-distortion performance is another crucial issue in DNN-based image compression.
The methods in~\cite{toderici2015variable,toderici2016full,johnston2017improved} treat entropy coding as a post-processing step.
Ball{\'e} \etal~\cite{balle2016end} explicitly formulated DNN-based image compression under the framework of rate-distortion optimization. Assuming $\bm y$ is statistically factorized, they learned piece-wise linear density functions to compute  differential entropy as an approximation to discrete entropy. In a subsequent work~\cite{balle2018variational}, each $y_i$ is assumed to follow zero-mean Gaussian with its own variance separately predicted using side information. %by supposing that all the codes follows a zero mean Gaussian distribution controlled by the scale hyperprior, some side information are learnt from the quantized codes to model the entropy loss.
Minnen \etal~\cite{minnen2018joint} combined the autoregressive and hierarchical priors, leading to improved rate-distortion performance.
Theis \etal~\cite{theis2017lossy} introduced a continuous upper bound of the discrete entropy with a Gaussian scale mixture.
%
% Li \etal~\cite{li2017learning} learned a content-based importance map to capture spatially informative image regions, based on which spatially adaptive bit allocation can be naturally achieved.
% The sum of the importance map is adopted as the ratio loss to train a encoder-decoder network.
%By extract a small cuboid around the code from the code maps as the context, a small convolutional network is proposed for post context-based entropy coding.
%
Rippel \etal~\cite{rippel2017real} described  pyramid-based  analysis and synthesis transforms with  adaptive code length regularization for real-time image compression.  An adversarial loss~\cite{goodfellow2014generative} is incorporated  to generate visually realistic results at low bit rates~\cite{rippel2017real}.
% Mentzer \etal~\cite{mentzer2018conditional1} proposed a masked CNN for entropy modeling, and optimized it by alternating between  the entropy model and the autoencoder optimization.
% Agustsson \etal~\cite{agustsson2017soft} introduced a soft-to-hard relaxation scheme by approximating  quantization  with a parametric softmax function.

\section{CCNs for Entropy Modeling}
% In data compression frameworks, the quantized codes still need to be compressed into bit stream by entropy coding.
% %
% As the key issue of entropy coding, a better estimation of the entropy can improve the compression ratio by a lot.
% %
% The very basic way is to suppose the codes are i.i.d and build a histogram of the codes to model the entropy which is widely used in many data compression frameworks.
% %
% However, the codes should be not independent and they are depending on certain context.
% %
% With a given coding order, all the codes coded before a given code $c$ inner the whole code block $\mathbf{C}$ is its coding context $\mbox{CTX}(c,\mathbf{C})$.
% %
% In context-based adaptive binary arithmetic coding~\cite{marpe2003context}, the context is proved to be very useful in entropy coding for image compression.
% %
% However, considering that the complexity of modeling large context increase exponentially with respect to the size of the context, \cite{marpe2003context} only take use of two codes in the context.
% %

% Recent progresses of the deep learning, like RNN, LSTM, PixelRNN and PixelCNN, have shown great power in modeling long range dependency in natural language processing and image generation, which throw light on modeling much large coding context in entropy modeling.
% %
% Considering the inefficiency of the RNN models, we in this paper proposed a modified fully convolutional network, slope convolutional network, for predicting the entropy of the codes from the code context.
%

\begin{figure}[!tbp]
	\centering
	\includegraphics[width=0.76\linewidth]{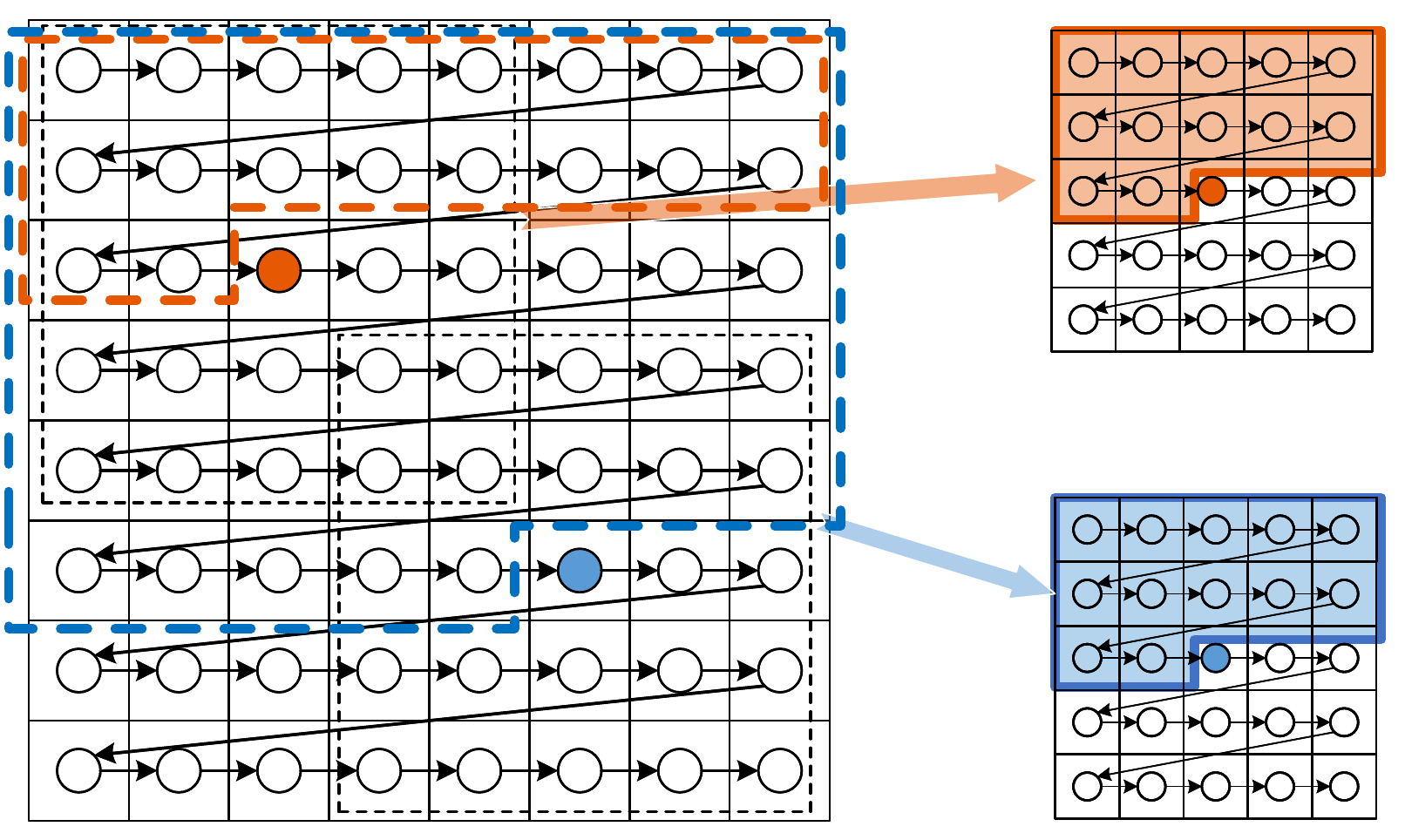}
	\caption{Illustration of 2D masked convolution in the input layer of the proposed CCN for entropy modeling. A raster coding order (left to right, top to bottom) and a convolution kernel size of $5\times 5$ are assumed here. The orange and blue dashed regions indicate the full context of the orange and blue codes, respectively. In the right panel, we highlight the support sets of the two codes in corresponding colors, which share the same mask.}
	\label{fig:conv}
\end{figure}

In this section, we present in detail the construction of CCNs for entropy modeling. We work with a fully convolutional network, consisting of $T$ layers of convolutions followed by point-wise nonlinear activation functions, and assume the standard raster coding order (see Fig.~\ref{fig:conv}). In order to perform efficient context-based entropy coding, two assumptions are made on the network architecture:
\begin{itemize}
\item For a code block $\bm y\in\mathbb{Q}^{M\times H\times W}$, where $M$, $H$, and $W$ denote the dimensions along channel, height, and width directions, respectively, the corresponding output of the $t$-th convolution layer $\bm v^{(t)}$ has a size of ${M\times H\times W\times N_t}$, where $N_t$ denotes the number of feature blocks to represent $\bm y$.
\item Let $\mathrm{CTX}( y_i(p,q),\bm y)$ be  the set of codes encoded before $y_i(p,q)$ (\ie, full context), and $\mathrm{SS}( v^{(t)}_{i,j}(p,q))$ be the set of codes in the receptive field of $v^{(t)}_{i,j}(p,q)$ that contributes to its computation (\ie, support set), respectively. Then, $\mathrm{SS}(v^{(t)}_{i,j}(p,q))\subset \mathrm{CTX}( y_{i}(p,q), \bm y)$. %In other words, given $\mathrm{CTX}(y_{i}(p,q), \bm y)$, $v^{(t)}_{i,j}(p,q)$ is completely determined with a fixed CCN.
% \item For  $\bm y'\subset \bm y$ and $y_i(p,q)\in\bm y'$, $\mathrm{CTX}(y_i(p,q),\bm y')\subset \mathrm{CTX}(y_i(p,q),\bm y)$.
\end{itemize}
Assumption I establishes a one-to-many correspondence between the input code block $\bm y$ and the output feature representation $\bm v^{(T)}$. In other words, the feature $ v^{(t)}_{i,j}(p,q)$ in $i$-th channel and $j$-th feature block at spatial location $(p,q)$ is uniquely associated with $y_{i}(p,q)$.
Assumption II ensures that the computation of $v^{(t)}_i(p,q)$ depends only on a subset of $\mathrm{CTX}(y_i(p,q),\bm y)$. Together, the two assumptions guarantee the legitimacy of context-based entropy modeling in fully convolutional networks, which can be achieved by placing translation-invariant binary masks to
convolution filters.

We start with the case of a 2D code block, where $\bm y\in \mathbb{Q}^{H\times W}$, and define masked convolution at the $t$-th layer as
% \begin{equation}
% 	F^{t+1,s}_{p,q} = \sum_{z=0}^{g_t-1}\left(\sum_{l-i=p,m-j=q}\mathbf{m}_{i,j}^t F^{t,z}_{l,m} w^{t,s,z}_{i,j}\right),
% 	\label{eq:conv2d}
% \end{equation}
\begin{equation}
 v^{(t)}_i(p,q) = \sum_{j=1}^{N_t} \left( u^{(t)}_j \ast\left( m^{(t)}  \odot w^{(t)}_{i,j}\right) \right)(p,q)+b_i^{(t)},
\label{eq:conv2d}
\end{equation}
{
where $\ast$ and $\odot$ denote 2D convolution and Hadamard product, respectively. $\bm w^{(t)}_{ij}$ is a 2D convolution filter, $\bm m^{(t)}$ is the corresponding 2D  binary mask, and $b_i^{(t)}$ is the bias.
}
According to Assumption I, the input $\bm u^{(t)}_i$  and the output $\bm v^{(t)}_i$ are of the same size as $\bm y$.
The input code block $\bm y$ corresponds to $\bm u^{(0)}_0$.

\begin{figure*}[!htp]
	\centering
	\begin{minipage}{1.0\linewidth}
     \includegraphics[width=1.0\textwidth]{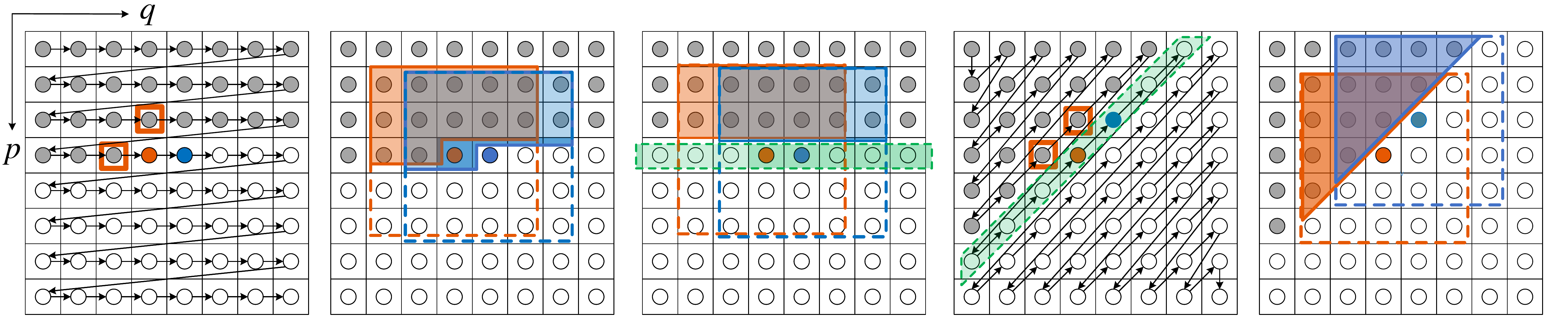}	
	\end{minipage}
	\begin{minipage}{1.0\linewidth}
		\begin{minipage}{0.205\linewidth} \centering{\scriptsize{ (a)}} \end{minipage}
		\begin{minipage}{0.175\linewidth} \centering{\scriptsize{(b)}} \end{minipage}	
		\begin{minipage}{0.21\linewidth} \centering{\scriptsize{(c)}} \end{minipage}
		\begin{minipage}{0.175\linewidth} \centering{\scriptsize{(d)}} \end{minipage}
		\begin{minipage}{0.2\linewidth} \centering{\scriptsize{(e)}} \end{minipage}
	\end{minipage}
	
	\caption{Illustration of  code dividing techniques in conjunction with different coding orders for a 2D code block. The orange and blue dots represent two neighbouring codes. The gray dots denote  codes that have already been encoded, while the white circles represent  codes yet to be encoded. (a) Raster coding order adopted in many compression methods. (b) Support sets of the orange and blue codes, respectively. It is clear that the orange code is in the support set of the blue one, and therefore should be decoded first. (c) Code dividing scheme for the raster coding order. By removing the dependencies among codes in each row, the orange and blue codes can be decoded in parallel. However, the orange code is excluded from the support set of the blue one, which may hinder  entropy estimation accuracy. (d) Zigzag coding order and its corresponding code dividing scheme. The two codes in the orange squares that are  important for the orange code in entropy prediction are retained in its partial context. (e) Support sets of the orange and blue codes in compliance with the zigzag coding order.}
	\label{fig:context}
\end{figure*}

\begin{figure*}[!tbp]
	\centering
	\begin{minipage}[b]{0.48\textwidth}
		\includegraphics[width=1.0\linewidth]{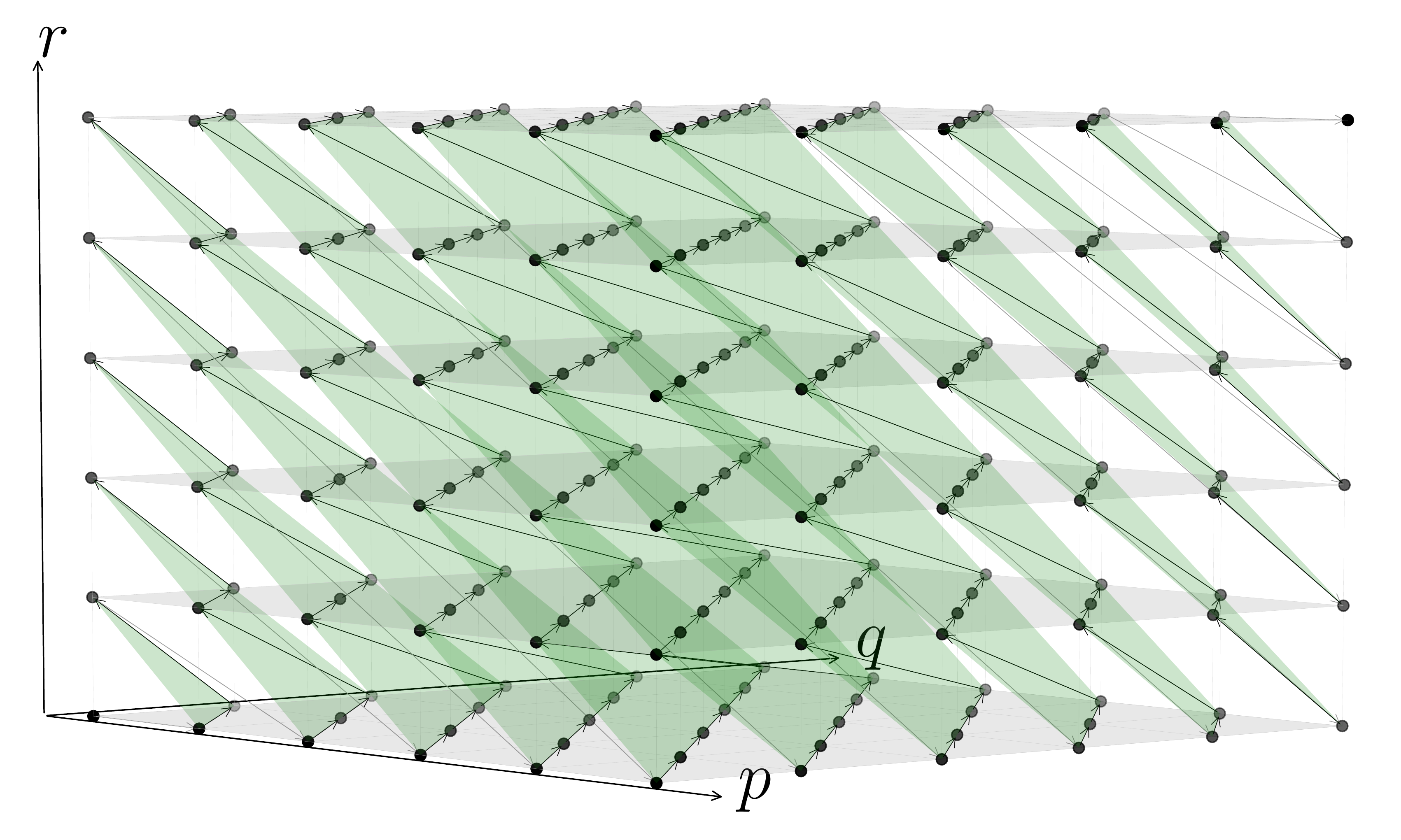}
		\centering{\scriptsize{(a)}}
	\end{minipage}
	\begin{minipage}[b]{0.48\textwidth}
		\includegraphics[width=1.0\linewidth]{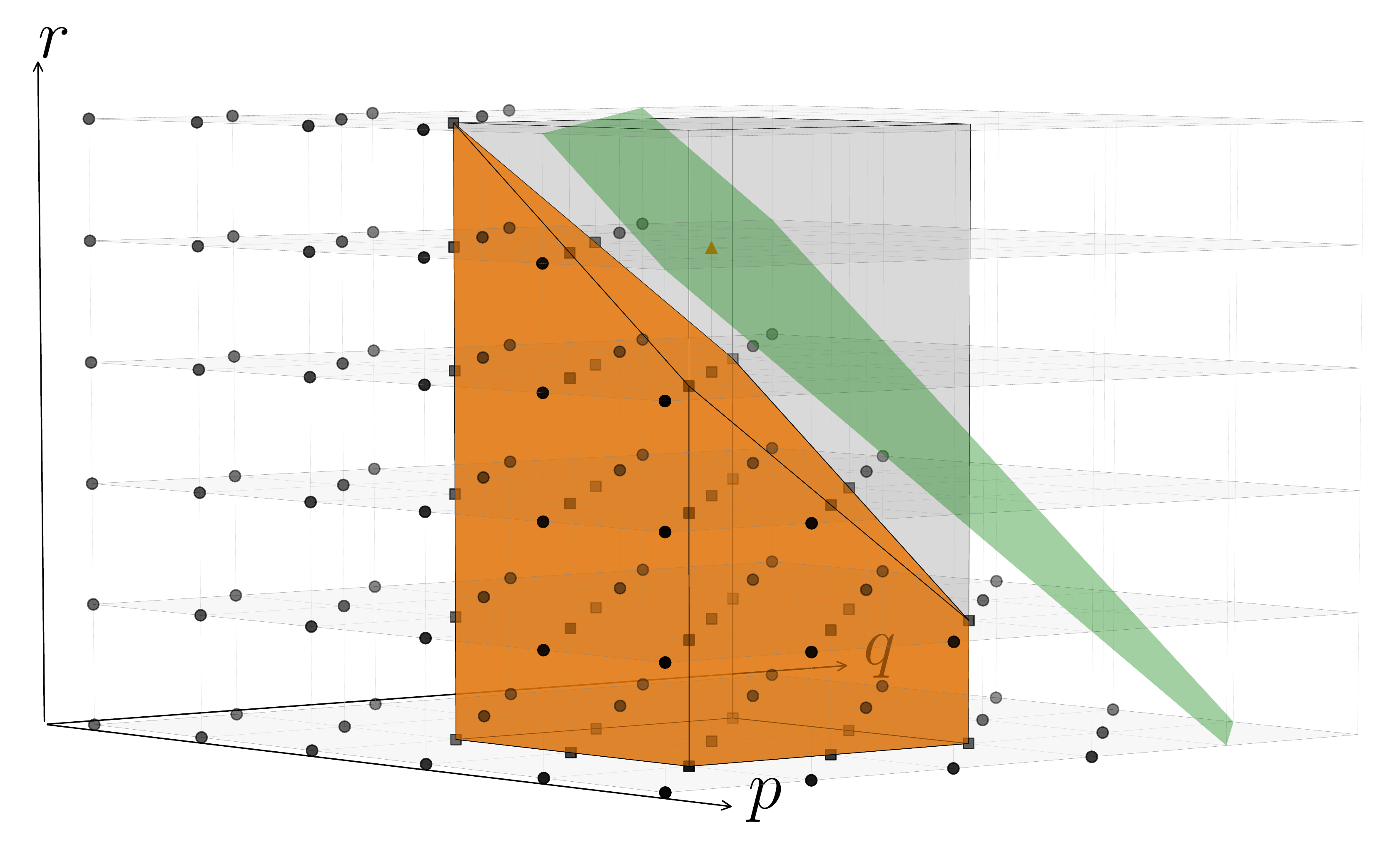}
		\centering{\scriptsize{(b)}}
	\end{minipage}
	\caption{Illustration of the proposed 3D zigzag coding order and 3D code dividing technique. (a) Each group in the shape of a diagonal plane is highlighted in green. Specifically, $\GRP_k(\bm y) =\{y_r(p,q)|r+p+q=k\}$ are encoded earlier than $\GRP_{k+1}(\bm y)$. Within $\GRP_k(\bm y)$, we first process codes along the line $p+q =k$ by gradually decreasing $p$. We then process codes along the line $p+q=k-1$ with the same order. The procedure continues until we sweep codes along the last line $p+q=\max(k-r,0)$ in $\GRP_{k}(\bm y)$. (b) Support set  of the orange code  with a spatial filter size  of $3\times 3$. Zoom in for improved visibility.}
	\label{fig:3d}
\end{figure*}

For the input layer of a fully convolutional network, the codes to produce $v^{(0)}_{i}(p,q)$ is
$\Omega_{p,q}=\{y(p+\mu,q+\nu)\}$, where $(\mu,\nu)\in\Psi$ is the set of local indices centered at $(0,0)$. We choose
\begin{align}
\mathrm{SS}(v^{(0)}_{i}(p,q))=\mathrm{CTX}(y(p,q),\Omega_{p,q})\subset \mathrm{CTX}(y(p,q),\bm y),
\end{align}
which can be achieved by setting
\begin{align}
	m^{(0)}{(\mu,\nu)} = \begin{cases}
	1, &\mbox{if } \Omega_{p,q}(\mu,\nu) \in\mathrm{CTX}(y(p,q),\bm y) \\
	0, &\mbox{otherwise}.
\end{cases}
	\label{eq:mask_input}
\end{align}
Fig.~\ref{fig:conv} illustrates the concepts of full context $\mathrm{CTX}(y(p,q), \bm y)$, support set $\mathrm{SS}(v^{(0)}(p,q))$, and  translation-invariant mask $\mathbf m^{(0)}$, respectively. At the $t$-th layer, %, the input to the convolution is the representation $\mathbf{U}^{(t)}$. The features to produce $\mathbf{V}^{(t)}_i{(p,q)}$ in $\mathbf{U}^{(t)}_j$ is $\mathbf{Q} ^{(t)}_j=\{\mathbf{U}^{(t)}_j(p+u,q+v)\}_{(u,v)\in \Omega}$. For
if we let $\bm m^{(t)} = \bm m^{(0)}$, for a code $y(p+\mu,q+\nu)\in \mathrm{CTX}(y(p,q), \bm y)$, we have
\begin{align}
\mathrm{SS}(u^{(t)}_j(p+\mu,q+\nu))\subset& \CTX(y(p+\mu,q+\nu),\bm y)\nonumber\\
\subset& \CTX(y(p,q),\bm y),
\end{align}
where the first line follows by induction and the second line follows from the definition of context.  That is, as long as $y(p+\mu,q+\nu)$ is in the context of $y(p,q)$, we are able to
compute $v^{(t)}_i{(p,q)}$ from $u^{(t)}_j(p+\mu,q+\nu)$  without violating Assumption II. In addition, $u^{(t)}_j(p,q)$ for $t>0$ is also generated from $\CTX(y{(p,q)},\bm y)$, and can be used to compute $v^{(t)}_i{(p,q)}$. Therefore, we may modify the mask at the $t$-th layer
\begin{align}
	\mathbf{m}^{(t)}{(\mu,\nu)} = \begin{cases}
	\mathbf{m}^{(0)}(\mu,\nu), &\mbox{if } (\mu,\nu)\ne(0,0)\\
	1,  &\mbox{otherwise}.
\end{cases}
	\label{eq:mask_hidden}
\end{align}

\begin{figure}[!tbp]
	\centering
	\begin{minipage}[b]{0.48\linewidth}
		\includegraphics[width=1.0\textwidth]{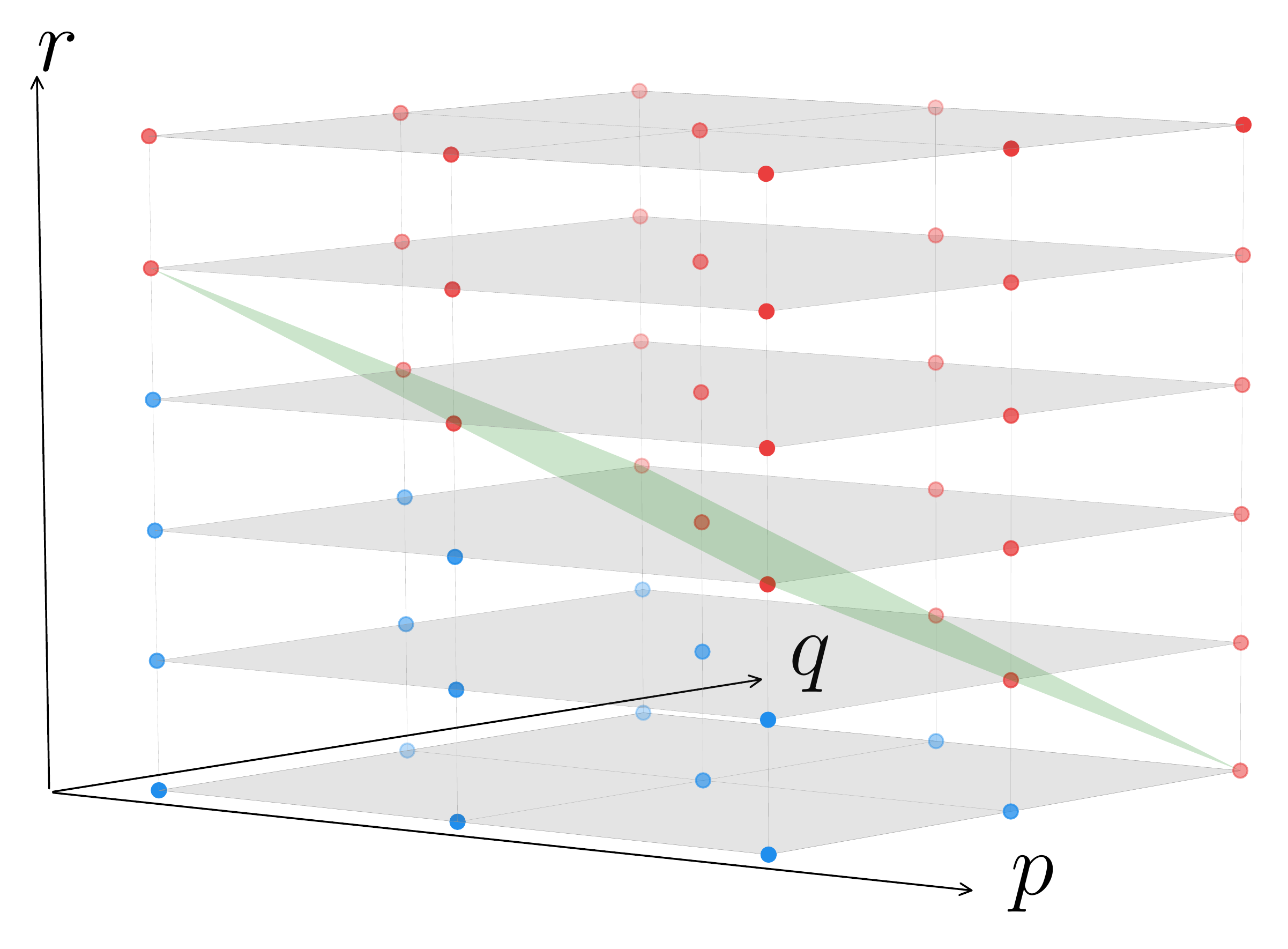}
		\centering{\scriptsize{(a)}}
	\end{minipage}
	\begin{minipage}[b]{0.48\linewidth}
		\includegraphics[width=1.0\textwidth]{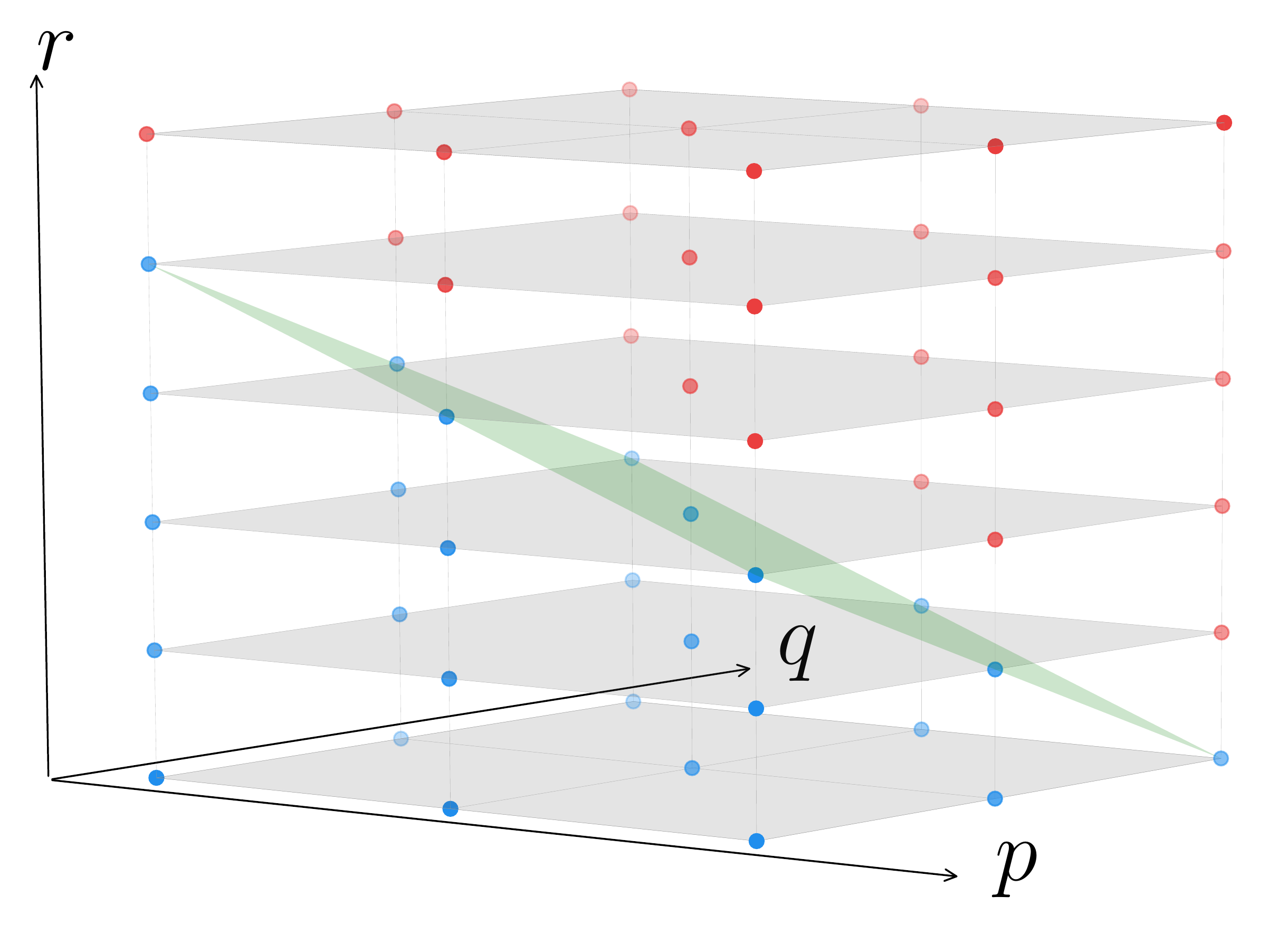}
		\centering{\scriptsize{(b)}}
	\end{minipage}
	\caption{Illustration of masked codes with $M=6$, $r=2$, and a filter size of $3\times 3$.  Blue dots represent codes activated by the mask and red dots indicate the opposite. The only difference lies in the green diagonal plane. (a) Input layer. (b) Hidden layer.}
	\label{fig:3d_mask}
\end{figure}

\begin{figure*}[!tbp]
	\centering

	\begin{picture}(500,200)
	\put(0,0){\includegraphics[width=1.0\textwidth]{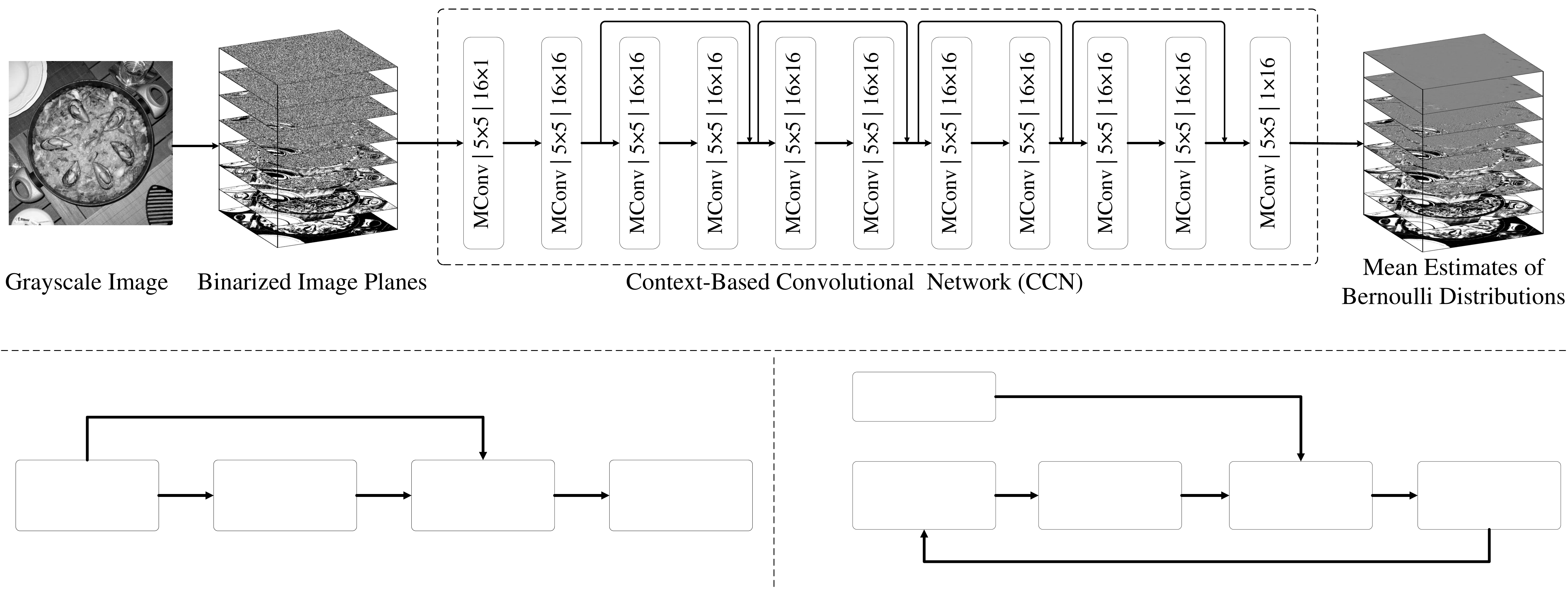}}
	\put (250,90){\scriptsize{(a)}}
	\put (120,-5){\scriptsize{(b)}}
	\put (380,-5){\scriptsize{(c)}}
	\put (30,16){\scriptsize{$\bm{y}$}}
	\put (118,50){\scriptsize{$P(\bm{y})$}}
	\put (90,16){\scriptsize{$P(\cdot)$}}
	\put (260,50){\scriptsize{$\GRP_0(\bm y),\ldots, \GRP_{k-1}(\bm y)$}}
	\put (372,50){\scriptsize{$P(\GRP_{k}(\bm y))$}}
	\put (478,50){\scriptsize{$\GRP_{k}(\bm y)$}}
	\put (9,32){\footnotesize{Code Block}}
	\put (86,32){\footnotesize{CCN}}
	\put (142,36){\footnotesize{Arithmetic}}
	\put (146,28){\footnotesize{Encoder}}
	\put (208,32){\footnotesize{Bitstream}}
	\put (290,36){\footnotesize{Decoded}}
	\put (285,28){\footnotesize{Code Block}}
	\put (356,32){\footnotesize{CCN}}
	\put (412,36){\footnotesize{Arithmetic}}
	\put (416,28){\footnotesize{Decoder}}
	\put (475,36){\footnotesize{Decoded}}
	\put (470,28){\footnotesize{Code Group}}
	\put (290,65){\footnotesize{Bitstream}}
	\end{picture}

	\caption{\cmt{Proposed lossless image compression method. (a) gives the CCN-based entropy model for lossless image compression. The grayscale image $\bm x$ is first converted to bit-plane representation $\bm y$, which is fed to the network to predict the mean estimates of Bernoulli distributions $P(y_r(p,q)|\mathrm{SS}(v_r(p,q)))$. Each convolution layer is followed by a parametric ReLU nonlinearity, except for the last layer, where a sigmoid function is adopted. From the mean estimates, we find that for most significant bit-planes, our model makes more confident predictions closely approximating local image structures. For the least significant bit-planes, our model is less confident, producing mean estimates close to $0.5$. MConv: masked convolution used in our CCNs with filter size ($S\times S$) and the number of feature blocks (output$\times$input). (b) and (c) show the arithmetic encoding and decoding with the learned CCN, respectively.}}
	\label{fig:lossless_frame}
\end{figure*}

\subsection{Proposed Strategies for Parallel Entropy Decoding}
With the translation-invariant masks designed in Eqn.~(\ref{eq:mask_input}) and Eqn.~(\ref{eq:mask_hidden}), the proposed CCN can efficiently encode $\bm y$ in parallel.
However, it remains difficult to parallelize the computation in entropy decoding.
As shown in Fig.~\ref{fig:context} (a) and (b), the two nearby codes in the same row (highlighted in orange and blue, respectively) cannot be decoded simultaneously  because the orange code is in the support set (or context) of the blue code given the raster coding order. To speed up entropy decoding, we may further remove  dependencies between codes at the risk of model accuracy. Specifically, we partition $\bm y$ into $K$ groups, namely, $\GRP_0(\bm y),\ldots, \GRP_{K-1}(\bm y)$, and assume the codes within the same group are statistically independent. This results in a partial context $\mathrm{PTX}(y(p,q),\bm y)=\{\GRP_0(\bm y),\ldots,\GRP_{k-1}(\bm y)\}$ for $y(p,q)\in \GRP_{k}(\bm y)$. In other words, all codes in the $k$-th group share the same partial context, and can be decoded in parallel. Note that code dividing schemes are largely constrained by pre-specified coding orders. For example, if we use a raster coding order, it is straightforward to divide $\bm y$ by row. In this case, $y(p,q-1)$ ($p$ and $q$ index vertical and  horizontal directions, respectively), which is extremely important in predicting the probability of $y(p,q)$ according to CABAC~\cite{marpe2003context}, has been excluded from its partial context. To make a good trade-off between modeling efficiency  and accuracy, we switch to a zigzag coding order as shown in Fig.~\ref{fig:context} (d), where $\GRP_k(\bm y)=\{y(p,q)| p+q=k\}$ and $\mathrm{PTX}(y(p,q),\bm y)=\{y(p',q')|p'+q'< k\}$. As such, we retain the most relevant codes in the partial context for better entropy modeling (see Fig.~\ref{fig:lossless_cmp2} for quantitative results). Accordingly, the mask at the $t$-th layer becomes
\begin{align}
	m^{(t)}{(\mu,\nu)} = \begin{cases}
	m^{(0)}(\mu,\nu), &\mbox{if } \mu+\nu\ne 0\\
	1,  &\mbox{otherwise}.
\end{cases}
	\label{eq:mask_hidden_slope}
\end{align}

% \begin{figure}[!tbp]
%     \centering
%     \includegraphics[width=0.7\linewidth]{figs/order.png}
%     \caption{The proposed coding order on a $3\times 3\times 3$ code block.}
%     \label{fig:my_label}
% \end{figure}

Now, we extend our discussion to a 3D code block, where  $\bm y\in \mathbb{Q}^{M\times H \times W}$.
Fig.~\ref{fig:3d} (a) shows the proposed 3D zigzag coding order and 3D code dividing technique. Specifically, $\bm y$ is divided into $K=M+H+W-2$ groups in the shape of diagonal planes, where the $k$-th group is specified by $\GRP_k(\bm y)=\{y_r(p,q)|r+p+q=k\}$. The partial context of $y_r(p,q)\in\GRP_k(\bm y)$ is defined as $\mathrm{PTX}(y_r(p,q),\bm y)=\{y_{r'}(p', q')|r'+p'+q'<k\}$. We then write masked convolution in the 3D case as
\begin{align}
v^{(t)}_{i,r}(p,q) = \sum_{j=1}^{N_t} \sum_{s=1}^{M}\left( u^{(t)}_{j,s} \ast\left( m^{(t)}_{r,s} \odot w^{(t)}_{i,j,r,s}\right) \right)(p,q)+b_{i,r}^{(t)},
\end{align}
where $\{i,j\}$ and $\{r,s\}$ are indices for the feature block and the channel, respectively.
For the 2D  case, each layer shares the same mask ($M=1$). When extending to 3D code blocks, each channel in a layer shares a mask, and there are a total of $M$ 3D masks. For the input layer, the codes to produce $v^{(0)}_{i,r}(p,q)$ is  $\Omega_{p,q}=\{y_s{(p+\mu,q+\nu)}\}_{(\mu,\nu)\in\Psi, 0\le s< M}$, based on which we define the mask as
\begin{align}
	m^{(0)}_{r,s}{(\mu,\nu)} = \begin{cases}
	1, &\mbox{if } \Omega_{p,q}(\mu,\nu) \in\mathrm{PTX}({y_r(p,q),\bm y})\\
	0, &\mbox{otherwise}.
\end{cases}
	\label{eq:3dmask_input}
\end{align}
For the $t$-th layer, we  modify the mask to include the current diagonal plane
\begin{align}
	m^{(t)}_{r,s}(\mu,\nu) = \begin{cases}
	m^{(0)}_{r,s}(\mu,\nu), &\mbox{if } s+\mu+\nu\ne r\\
	1, &\mbox{otherwise},
	\end{cases}
	\label{eq:mask_input_3d}
\end{align}
as shown in Fig.~\ref{fig:3d_mask}.

\subsection{Analysis of the Degree of Parallelism}\label{sec:dop}
{
We analyze the degree of parallelism (DOP) of the proposed CCN, which is defined as the number of operations that can be or are being simultaneously executed by a computer. For entropy encoding, all codes can be  estimated in parallel, leading to  a DOP of $MHW$.
%
%For entropy decoding, DNN-based auto-regressive models~\cite{mentzer2018conditional1,oord2016pixel,salimans2017PixeCNN} need to process one code at a time with a DOP of one. 
The proposed CCN divides the code block into $K=M+H+W-2$ groups and allows the codes within each group to be processed in parallel. Ideally, the DOP should be the group size (\textit{i.e.}, the number of codes in the group). However, different groups have different sizes in the shape of irregular diagonal planes. Here, the average group size is adopted as a rough indication of DOP,
\begin{align}
    \mathrm{DOP}=\frac{1}{K}\sum_{k=0}^{K-1} Z_k=\frac{MHW}{M+H+W-2},
\end{align}
where $Z_k$ denotes the number of codes in the $k$-th group.
It is clear that DOP is a function of height,  width, and  channel of the  code block.}

\section{CCN-Based Entropy Models for Lossless Image Compression}
 In this section, we combine our CCN-based entropy model with the arithmetic coding algorithm for lossless image compression.

\begin{figure*}[!tbp]
	\centering
	\begin{picture}(500,180)
\put(0,0){\includegraphics[width=1.0\textwidth]{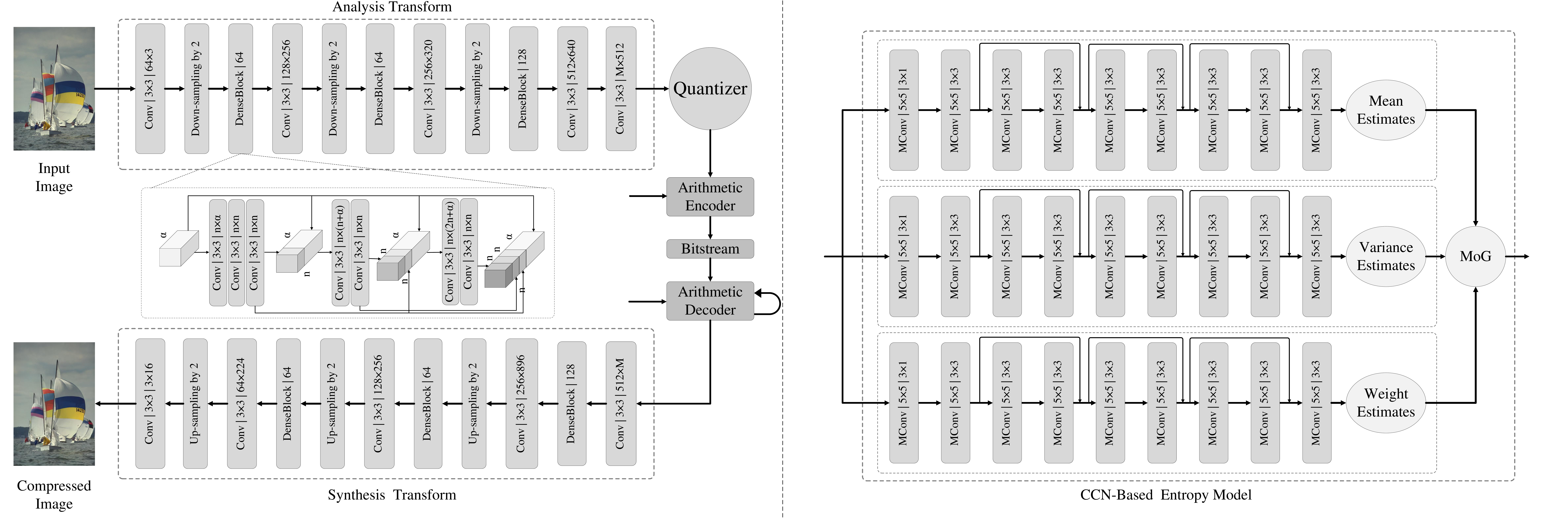}}
\put (16,100){\scriptsize{$\bm{x}$}}
\put (213,146){\scriptsize{$\bm{z}$}}
\put (236,120){\scriptsize{$\bm{y}$}}
\put (262,86){\scriptsize{$\bm{y}$}}
\put (505,85){\scriptsize{$P(\bm{y})$}}
\put (187,105){\scriptsize{$P(\bm{y})$}}
\put (187,70){\scriptsize{$P(.)$}}
\put (236,50){\scriptsize{$\bm{y}$}}
\end{picture}
	\caption{The architecture of the proposed  lossy image compression method, which consists of an analysis transform $g_a$, a non-uniform and trainable quantizer $g_d$, a CCN-based entropy model, and a synthesis transform $g_s$. Conv: regular convolution with filter support ($S\times S$) and number of channels (output$\times$input). Down-sampling: implemented jointly with the adjacent convolution (also referred to as stride convolution). DenseBlock: $m$ matches the input channel number of the preceding convolution. $n$ is the channel number in DenseBlock.  MConv: masked convolution used in our CCNs with filter size ($S\times S$) and the number of feature blocks (output$\times$input). Note that the number of channels is fixed in MConv, and is determined by that of ${\bm y}$. \cmt{More details about the arithmetic encoding and decoding can be found in Fig.~\ref{fig:lossless_frame}(b)(c).}}
	\label{fig:lossy_framework}
\end{figure*}
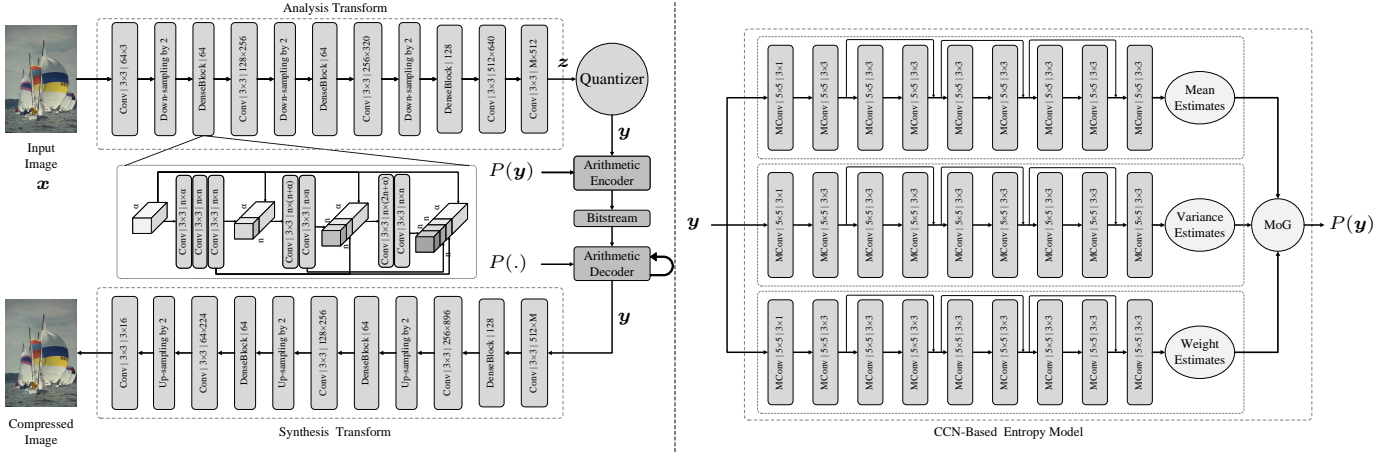

As a starting point, we  binarize the grayscale image $\bm x\in \mathbb{R}^{H\times W}$ to obtain a 3D code block
\begin{align}
y_{r}(p,q) = \left\lfloor \frac{x(p,q)}{2^{7-r}} \right\rfloor {\rm mod}\ 2, \quad r=0,1,\ldots,7,
\end{align}
where we index the most significant bit-plane with $r=0$.
% As proved in ~\cite{said2004introduction}, arithmetic coding~\cite{witten1987arithmetic} (AC) is the optimal coding scheme to approximating the code length calculated by the predicted entropy. It compress the whole code string (serialized from the code block with the given coding order) into a single number in the interval of $[0,1]$. The arithmetic coding keeps an interval with a start point and an end point to mark the interval in the whole coding process.  Given $S=\{s_0,\ldots,s_{m-1}\}$ the set of different symbols in the code string and a code to be coded $c\in \mathbf{C}$,  the AC first predict a discrete distribution $P(c=s_i)$ with $i=0,\ldots,m-1$ for $c$ and then divide the current interval into $m$ different sub-intervals according to the distribution.
% %
% After that, with $s_t = c$, the AC updates the current interval to the $t$-th sub-interval.
% %
% Starting from the interval $[0,1]$ and repeating the process above until all the codes in the code string have been processed,  the AC finally stop with a interval and a single value inner the stop interval can be used to represent the whole code string.
% With the description of arithmetic coding, the most important thing in AC is to predict the discrete distribution of each code. Given the discrete distortion, the AC can compress the codes into a bit stream with the length of $-\sum_{c\in\mathbf{C}}\log_2(P(c))$. With a better estimation of the discrete distribution, the codes can be compressed into a smaller size. Thus, we adopt the introduced SCN to predict the discrete distribution from the code context.
Our CCN takes $\bm y$ as input and produces a feature block $\bm v$ (the superscript $(T)$ is omitted for notation convenience) of the same size to compute the mean estimates of Bernoulli distributions  $P(y_r(p,q)|\mathrm{SS}(v_r(p,q)))$. Fig.~\ref{fig:lossless_frame} shows the network architecture, which has eleven masked convolution layers with parametric ReLU nonlinearities in between. The last convolution responses undergo a sigmoid nonlinearity to constrain their range between $[0,1]$. We make four residual connections as suggested in~\cite{he2016deep} to accelerate training. We will experiment with two hyper-parameters in CCN: the convolution filter size for all layers   $S$ and the number of feature blocks in hidden layers $N$.
%All slope convolutions adopt $z\times z$ kernels to produce $N$ feature blocks and PReLU nonliearity except for the last slope convolution layer which adopt $2$ groups corresponding to the number of different symbols in $\mathbf{C}$ and a softmax nonlinearity for generating discrete probability table.

To optimize the network parameters collectively denoted by $\bm \theta$, we adopt the expected code length as the empirical loss
\begin{align}
	\ell(\bm \theta) = \mathbb{E}_{\bm y}&\Big[-\sum_{r,p,q}  \big({\mathbbm 1}(y_r(p,q)=1)\log_2( v_{r}(p,q))\nonumber\\
	&- \mathbbm{1}(y_r(p,q)=0)\log_2(1- v_{r}(p,q)) \big)\Big],
	\label{eq:lossless_entropy}
\end{align}
where $\mathbbm{1}(\cdot)$ is an indicator function and the expectation may be approximated by averaging over a mini-batch of training images. Finally, we implement our own arithmetic coding method with the learned CCN-based entropy model to compress $\bm y$ to bitstreams, and report performance using actual bit rates. %This facilitates comparison against widely used image compression standards.

\section{CCN-Based Entropy Models for Lossy Image Compression}
In lossy image compression, our objective is to minimize a weighted sum of rate and distortion, $\ell_r + \lambda \ell_d$, where $\lambda$ governs the trade-off between the two terms. As illustrated in Fig.~\ref{fig:lossy_framework}, the proposed compression method has four components: an analysis transform $g_a$, a quantizer $g_d$, a CCN-based entropy model, and a synthesis transform $g_s$. The analysis transform $g_a$ takes a color image $\bm x$ as input and produces the latent code representation $\bm z$, which is further quantized to generate the discrete code block $\bm y$. $g_a$ consists of three convolutions, each of which is followed by down-sampling with a factor of two. %The number of convolution filters in the three layers are set to $64$, $128$, and $256$, respectively.
A dense block~\cite{huang2017densely} comprised of seven convolutions is employed after each down-sampling. After the last dense block, we add another convolution layer with $M$ filters to produce $\bm z$. Empirically,  the parameter $M$ sets the upper bound of the bit rate that a general DNN-based compression method can achieve. The parameters of $g_a$ constitute the parameter vector $\bm \phi$ to be optimized.

The synthesis transform $g_s$ has a mirror structure of the analysis transform. Particularly, the depth-to-space reshaping~\cite{toderici2016full,shi_2016_real} is adopted to up-sample the feature maps. The last convolution layer with three filters is responsible for producing the compressed  image in RGB space. The parameters of $g_s$ constitute the parameter vector $\bm \psi$ to be optimized.

For the quantizer $g_d$, we  parameterize its quantization centers for the $r$-th channel by $\{\omega_{r,0},\ldots,\omega_{r,L-1}\}$, where $L$ is the number of quantization centers and $\omega_{r,0}\le\ldots\le\omega_{r,L-1}$. The monotonicity of $\bm \omega$ can be enforced by a simple  re-parameterization based on cumulative functions. Given a fixed set of $\bm \omega$, we perform quantization by mapping $z_r(p,q)$ to its nearest  center that minimizes the quantization error
\begin{align}
y_r(p,q) = g_d(z_r(p,q))= \argmin_{\{ \omega_{r,l}\}}\Vert z_{r}(p,q) - \omega_{r,l} \Vert^2_2.
	\label{eq:mapping}
\end{align}
$g_d$ has zero gradients almost everywhere, which hinders training via back-propagation.  Taking inspirations from binarized neural networks~\cite{courbariaux2016binarized,zhou2016dorefa,rastegari2016xnor}, we make use of an identify mapping  $\hat{g}_d(y_r(p,q)) = y_r(p,q)$ as a continuous proxy to the step quantization function. During training, we use $g_d$ and $\hat{g}_d$ in the forward and backward passes, respectively.

The quantization centers $\bm \omega$ should be optimized by minimizing the mean squared error (MSE),
\begin{align}
	\ell(\bm \omega)=\frac{1}{MHW} \sum_{r,p,q}\Vert z_{r}(p,q)-
y_{r}(p,q))\Vert^2_2,
	\label{eq:loss_quant}
\end{align}
which is essentially a $k$-means clustering problem, and can be solved efficiently by the Lloyd's algorithm~\cite{lloyd1982least}. Specifically, we initialize $\bm \omega$ using uniform quantization, which appears to work well in all experiments. To make parameter learning of the entire model smoother, we adjust $\bm \omega$ using stochastic gradient descent instead of a closed-form update.

Without prior knowledge of the  categorical distributions of the quantized codes ${\bm y}$, we choose to work with   discretized MoG distributions, whose parameters are predicted by the proposed CCNs.  We write the differentiable MoG distribution with $C$ components as
\begin{align}
y_r(p,q) \sim \sum_{i=0}^{C-1}\pi_i\mathcal{N}(y_r(p,q);\mu_i, \sigma^2_i),
	\label{eq:mog}
\end{align}
where $\pi_i$, $\mu_i$ and $\delta^2_i$ are the mixture weight, mean, and variance of the $i$-th component, respectively. Then, {the probability of $y_r(p,q)$ is calculated as the integral over the quantization bin $\Delta$ that the code lies in}
\begin{align}
	P(y_{r}{(p,q)})=\int_{\Delta}\sum_{i=0}^{C-1}\pi_i\mathcal{N}(\xi; \mu_i, \sigma_i^2)d\xi.
	\label{eq:prob}
\end{align}
{For example, suppose $y_r(p,q)=\omega_{r,l}$, then $\Delta=[(\omega_{r,l-1}+\omega_{r,l})/2,(\omega_{r,l}+\omega_{r,l+1})/2]$, where we define $\omega_{r,-1} = -\infty$ and $\omega_{r,L} = \infty$, respectively.}

% Since $G(\mathbf{T}_{r}{(p,q)})$ is continuous and derivable, $P(\hat{\mathbf{e}}_{r}{(p,q)})$ is also derivable with respect to $\hat{\mathbf{e}}_{r}{(p,q)}$ and the derivative is:
% \begin{equation}
% 	\frac{\partial P(\hat{\mathbf{e}}_{r}{(p,q)})}{\partial \hat{\mathbf{e}}_{r}{(p,q)}} = G(\mathbf{T}_{r}{(p,q)})|_{\frac{z_{r,t}+z_{r,t-1}}{2}} - G(\mathbf{T}_{r}{(p,q)})|_{\frac{z_{r,t}+z_{r,t+1}}{2}}
% 	\label{eq:derivate}
% \end{equation}\textbf{}
%
Next, we describe the proposed entropy model in lossy image compression, which is comprised of three CCNs with the same structure, as shown in Fig.~\ref{fig:lossy_framework}. Each CCN consists of nine masked convolutions with three residual connections to produce $C$ feature blocks, matching the number of components in MoG. They separately output mixture weights, means and variances to build the discretized MoG distributions. The network parameters of our CCNs constitute the parameter vector $\bm \theta$ to be optimized.

\begin{figure}[!tbp]
	\centering
	\includegraphics[width=1.0\linewidth]{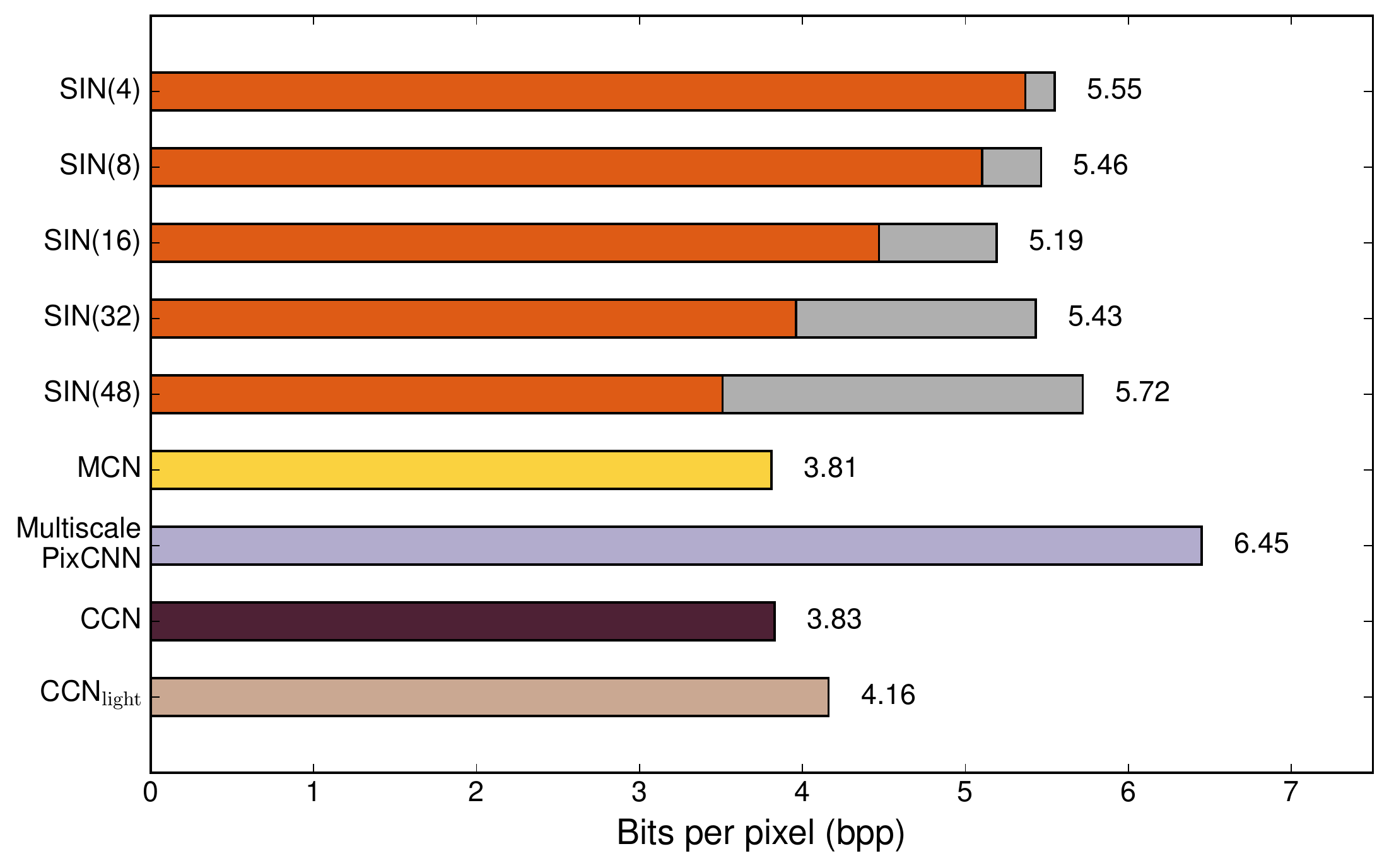}
	\caption{Bit rates (in terms of bpp) of different DNN-based entropy models for lossless image compression on the Kodak dataset. SIN($M$) refers to a side information network that allocates $M$ output channels to represent side information. The orange and gray bars represent the bit rates resulting from the image and the side information, respectively.}
	\label{fig:lossless_performance}
\end{figure}

Finally, we are able to write the empirical rate-distortion objective for the parameters $\{\bm \phi, \bm \psi, \bm \theta\}$ as
\begin{align}
\ell(\bm \phi, \bm \psi, \bm \theta)= \mathbb{E}_{\bm x}\Bigg[&-\sum_{i} \log_2 P_{y_i}\bigg(g_d\Big(g_a(\bm x;\bm \phi)\Big);\bm\theta\bigg) \nonumber\\
&+ \lambda \ell_d\bigg(g_s\Big(g_d\big(g_a(\bm x;\bm \phi)\big);\bm \psi\Big),\bm x\bigg)\Bigg].
	\label{eq:loss}
\end{align}
$\ell_d$ is the distortion term, which is more preferable to be assessed in perceptual space. In this paper, we optimize and evaluate our lossy image compression methods using the standard MSE and a perceptual metric MS-SSIM~\cite{wang2003multiscale}. Similar to lossless image compression, we combine the optimized entropy model with arithmetic coding and report the actual bit rates.

\section{Experiments}\label{sec:exp}
In this section, we test the proposed CCN-based entropy models in lossless and lossy image compression by comparing to state-of-the-art image coding standards and recent deep image compression algorithms. We first collect $10,000$ high-quality and high-definition images from Flickr, and  down-sample them to further reduce possibly  visible artifacts. We crop $1,280,000$ grayscale patches of size $128\times128$ and $640,000$ color patches of size  $3\times256\times256$ as the training sets for lossless and lossy image compression, respectively. We test our models on two independent datasets - Kodak and Tecnick~\cite{asuni2013testimages}, which are widely used to benchmark image compression performance. The Caffe implementations along with the pre-trained models are made available at \url{https://github.com/limuhit/CCN}.

\subsection{Lossless Image Compression}\label{sec:lossless_exp}
We train our CCN-based entropy models using the Adam stochastic optimization package~\cite{kingma2014adam} by minimizing the objective in Eqn.~(\ref{eq:lossless_entropy}). We start  with a learning rate of $10^{-4}$, and subsequently lower it by a factor of $10$ when the loss plateaus, until $10^{-6}$. The (actual) bit rate in terms of bits per pixel (bpp) is used to quantify the compression performance, which is defined as the ratio between the total amount of bits used to code the image and the number of pixels in the image. A smaller bpp indicates better performance. For example, an uncompressed grayscale image has eight bpp.

We first compare the proposed CCNs with masked convolutional networks (MCNs)~\cite{mentzer2018conditional1,li2018efficient}, Multiscale PixelCNN~\cite{reed2017multi}, and side information networks (SINs) ~\cite{balle2018variational} for entropy modeling.
As a special case of CCNs, MCNs specify the raster coding order without using any code dividing technique (see  Fig.~\ref{fig:context}). We implement our own version of MCN that inherits the network architecture from CCN with $N=16$ (number of feature blocks) and $S =5$ (filter size).
%Multiscale PixelCNN~\cite{salimans2017PixeCNN} is originally designed for image inpainting as a generative model. Here we adapt it for entropy modeling. Starting from a down-sampled grayscale image with a factor of four, we use a DNN of  similar model complexity compared with our CCN to predict the categorical distributions from the down-sampled image.
%
{
Multiscale PixelCNN~\cite{reed2017multi} is originally designed for image generation. Here we slightly modify it for entropy modeling. Specifically. we approximate the iterative process of Multiscale PixelCNN with a single network by mapping the initial grayscale image directly to the probability distributions of the original image. The network includes an up-sampling convolution layer followed by five residual blocks. We test various configurations of Multiscale PixelCNN, and choose to down-sample the original image by a factor of 4 as initialization, which delivers the best overall compression performance.
%PixelCNN++~\cite{salimans2017PixeCNN} is originally designed for image inpainting as a generative model. It starts from a smaller image, extends it 2 times larger in each step and repeats this until get the wanted image. Here, we adopt it for entropy modeling. Since the output of the entropy modeling are probability distributions instead of images, it is hard to directly adopts the PixelCNN++ for entropy prediction. We unfold the PixelCNN++ and directly approximate the process of PixelCNN++ with a single network by mapping the down-sampled image to the probability distributions. The used CNN network consists of an up-sampling convolution layer to up-sample the smaller image to the scale of the grayscale image followed by 5 residue blocks with each has two convolution layers. In Table~\ref{tab:NS}, we test PixelCNN++ on down-sampled grayscale images with the factor of 2,4 and 8. As we can see, larger start-up images will lead to smaller entropy. However, the bits used to save the start-up images also increase accordingly. Among the 3 models, the down-sampled grayscale image with a factor 4 get the best compression ratio. In comparison, we adopts the down-sampled grayscale image with a factor of four as the default setting for PixelCNN++.
}
SINs summarize the side information of ${\bm y}$ with a separate DNN, which is beneficial in probability estimation.  \cmt{We adopt a DNN-based autoencoder including three stride convolutions and two residual connections to generate the side information}, which is further quantized and compressed with arithmetic coding for  performance evaluation. We test five variants of SINs with different amount of side information by changing the number of output channels $M$.  All competing models are trained on the same dataset described at the beginning of Section~\ref{sec:exp}.

% \begin{table}[t]
%   \centering
%   \caption{Running time in seconds for different number of feature blocks ($N$) \& filter support ($S$) setting}\label{tab:NS}
%   \begin{tabular}{C{1.1cm} C{0.8cm} C{0.8cm} C{0.8cm} C{0.8cm} C{0.8cm} C{0.8cm}}
%     \toprule
%       ($N$,$S$) &  ($4$,$3$) &  ($4$,$5$)   &  ($8$,$3$) &   ($8$,$5$)   &  ($16$,$3$)& ($16$,$5$)\\
%     \midrule
%     Decoding time	& $6.69$	& $8.76$ &	$12.65$	& $19.62$ &	$30.89$ & $35.28$\\
%     \bottomrule
%   \end{tabular}
% \end{table}

% \begin{table}[t]
%   \centering
%   \caption{Running time in seconds for different patch size}\label{tab:patch}
%   \begin{tabular}{c c c c c c }
%     \toprule
%      $R$ & $1$ & $2$   & $4$ & $8$   & $16$\\
%     \midrule
%     Decoding time	& $6.69$	& $3.65$ &	$2.14$	& $1.45$ &	$0.983$\\
%     \bottomrule
%   \end{tabular}
% \end{table}

We also introduce a light-weight version of our method, which we name CCN$_\mathrm{light}$, by making the network architecture lighter (with $N=4$ and $S=3$) and by dividing the input image into non-overlapping patches for parallel processing.
{
%In detail, for an input image with the size of $h \times w$, we divide it into $t \times t$ smaller patches with the size of $\frac{h}{t}\times\frac{w}{t}$. The updated DOP should be $\frac{8hw}{8+h/t+w/t}$. Compared with the CCN without patch dividing strategy whose DOP is $\frac{8hw}{8+h+w}$, the patch dividing strategy should speed up CCN by about $t$ times. In CCN$_\mathrm{light}$, $t=16$. To further evaluate the effectiveness of lighter structure and the patch dividing strategy, we test CCN with different patch size and  number of feature blocks ($N$) \& filter support ($S$). In experiments, we fix the $N=4$ and $S=3$ and test the influence of different patch size controlled by $t$. Table~\ref{tab:patch} shows the running time with different patch size in decoding, i.e., $t=1,2,4,8,16$. In analysis, smaller patches, i.e., $t=16$, could speed up the decoding by nearly 16 times. However, in practice, the smaller patch size only speed up CCN by 6.8 times due to  our implementation and the device. For evaluating the effects of $N$ \& $S$, we fix $t=16$ and test CCN with different parameter settings. Table~\ref{tab:NS} shows the running time of CCN with different N and S, i.e., CCN($N$,$S$). Thus, smaller $N$ \& $S$ speeds up the CCN by 5.3 times. To summarize, smaller patch size contributes a little more than smaller $N$ \& $S$.
We split the image into a total of $R\times R$ smaller patches with the size of $\frac{H}{R}\times\frac{W}{R}$. The DOP now becomes $\frac{MHW}{M+H/R+W/R-2}$, which further speeds up the original CCN by about $R$ times.
%
%
%
% Specifically, we first fix $N=4$, $S=3$,  and test different patch sizes governed by $R$. Table~\ref{tab:patch} shows the running time with different patch sizes $R=\{1,2,4,8,16\}$ during decoding. As analysed above, smaller patches, \textit{e.g.}, $R=16$, could speed up the decoding by nearly $16$ times. In practice, we only observe an improvement of  $6.8$ times due to of the unoptimized implementation and the limitations of the device. For evaluating the effects of $N\&S$, we fix $R=16$ and test CCN with different parameter settings. From Table~\ref{tab:NS}, we find setting $N=4$ and $S=3$ speeds up the CCN by $5.3$ times compared to $N=16$ and $S=5$.
}

\begin{figure}[!tbp]
	\centering
	\includegraphics[width=1.0\linewidth]{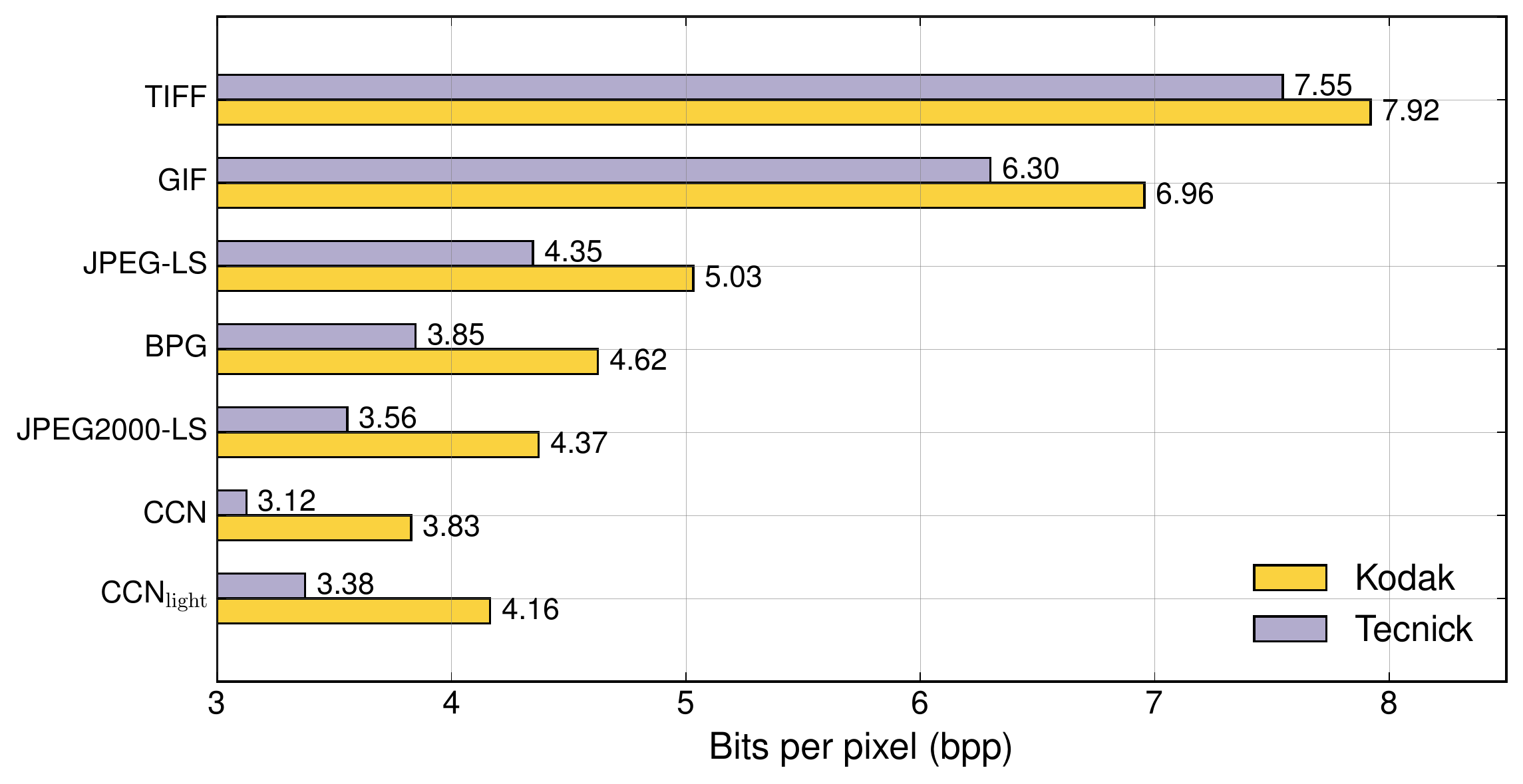}
	\caption{Bit rates of CCN in comparison with lossless image compression standards on the Kodak and Tecnick datasets.}
	\label{fig:lossless_cmp3}
\end{figure}

\begin{table}[!tbp]
	\scriptsize
	\caption{Running time in seconds and model complexity of different DNN-based entropy models on the Kodak dataset}
	\label{tab:lossless_time}

	\centering
	\begin{tabular}{cccccc}
		\toprule
		&SIN&MCN&\specialcell{Multiscale \\ PixelCNN}& CCN&$\mbox{CCN}_\mathrm{light}$     \\
		\midrule
		Encoding time& 0.155& 0.323& 0.400 & 0.323 & 0.074    \\
		Decoding time& 0.155& 3079.68& 0.400 & 35.28 & 0.984    \\
		\specialcell{\# of parameters \\ ($\times 10^6)$}& $6.70$& $3.89$&$3.67$&$3.89$&$0.10$\\
		\specialcell{\# of FLOPs \\ ($\times 10^{12})$}& $0.58$& $2.78$&$2.87$&$2.78$&$0.07$\\
		\bottomrule
		
	\end{tabular}
\end{table}

Fig.~\ref{fig:lossless_performance} shows the bit rates of the competing methods on the Kodak dataset. The proposed CCN matches the best performing model MCN, which suggests that with the proposed zigzag coding order and code dividing technique, CCN includes the most important codes as the partial context of the current code being processed. The bit rates of SINs come from two parts - the image itself and the side information. It is clear from the figure that increasing the amount of side information leads to bit savings of the image,  at the cost of additional bits introduced to code the side information. In general, it is difficult to determine the right amount of side information for optimal compression performance. Multiscale PixelCNN can also be regarded as a special case of SINs, whose side information is a small image without further processing, leading to the worst performance.

We also compare CCN  with the widely used lossless image compression standards, including TIFF, GIF, PNG, JPEG-LS, JPEG2000-LS, and BPG. All test results are generated by MATLAB2017. From Fig.~\ref{fig:lossless_cmp3}, we find that CCN (along with its light-weight version CCN$_\mathrm{light}$) overwhelms all competing methods on the Kodak/Tecnick dataset, achieving more than $5.9\%/6.2\%$ bit savings compared to the best lossless image compression standard, JPEG2000-LS.

\begin{figure}[!tbp]
	\centering
	\includegraphics[width=1.0\linewidth]{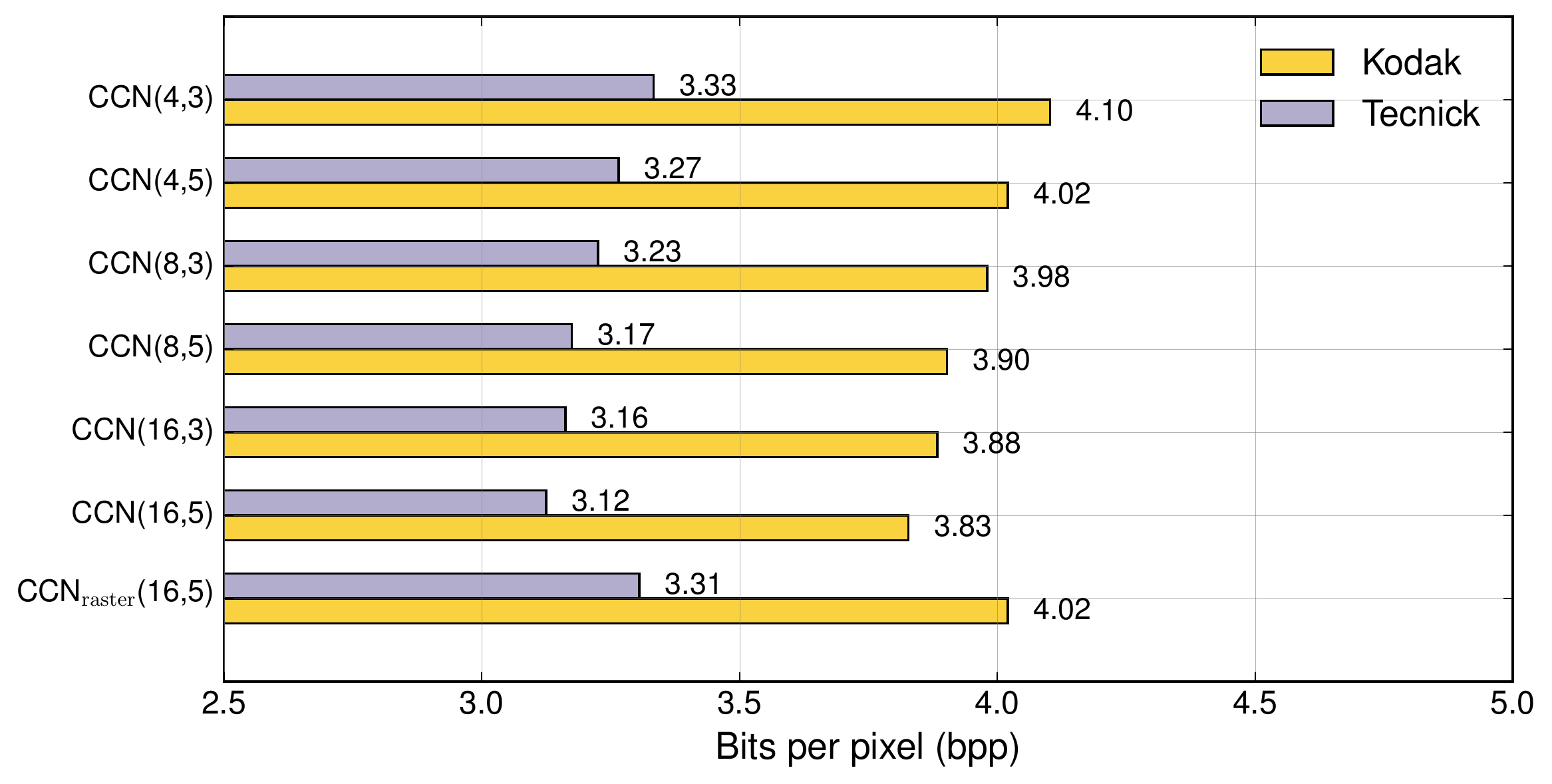}
	\caption{Ablation study of CCN on the Kodak and Tecnick datasets. CCN($N$,$S$) denotes the CCN with $N$ feature blocks and $S\times S$ filter size. CCN$_\mathrm{raster}$ represents the CCN with the raster coding order and the corresponding code dividing technique (see Fig.~\ref{fig:context}).}
	\label{fig:lossless_cmp2}
\end{figure}

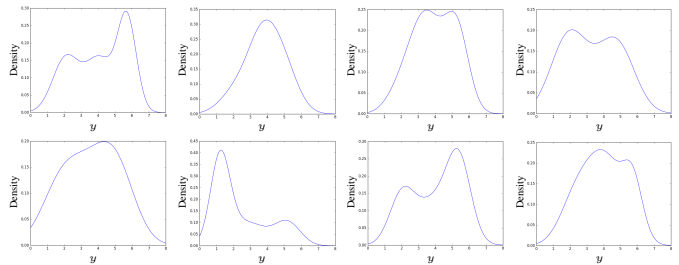
\begin{figure}[!tbp]
	\centering
	\begin{minipage}[t]{0.24\linewidth}
		\begin{picture}(50,50)
		\put(4,4){\includegraphics[width=0.92\linewidth]{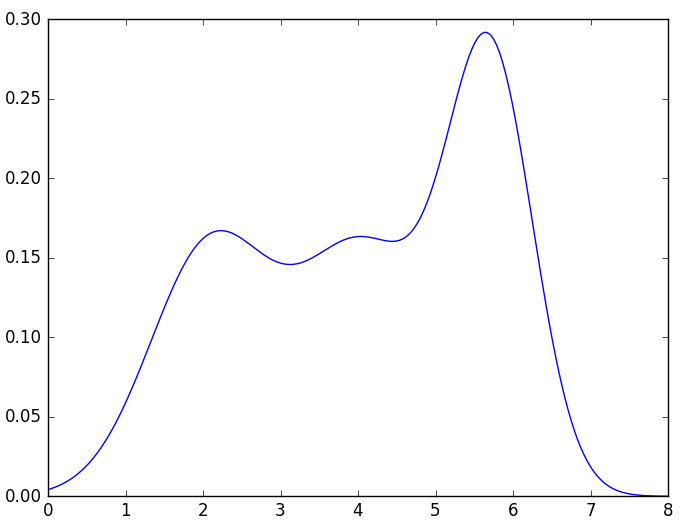}}
		\put(0,20){\rotatebox{90}{\mtiny{Density}}}
		\put(30,0){\mtiny{$y$}}
		\end{picture}
	\end{minipage}
	\begin{minipage}[t]{0.24\linewidth}
		\begin{picture}(50,50)
		\put(4,4){\includegraphics[width=0.92\linewidth]{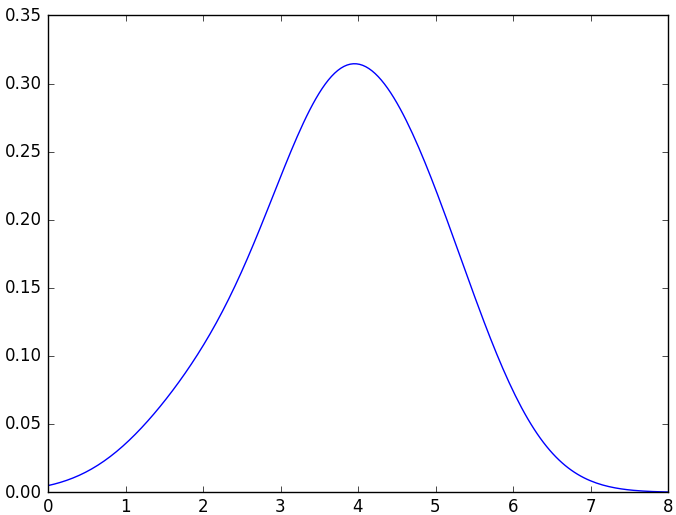}}
		\put(0,20){\rotatebox{90}{\mtiny{Density}}}
		\put(30,0){\mtiny{$y$}}
		\end{picture}
	\end{minipage}
	\begin{minipage}[t]{0.24\linewidth}
		\begin{picture}(50,50)
		\put(4,4){\includegraphics[width=0.92\linewidth]{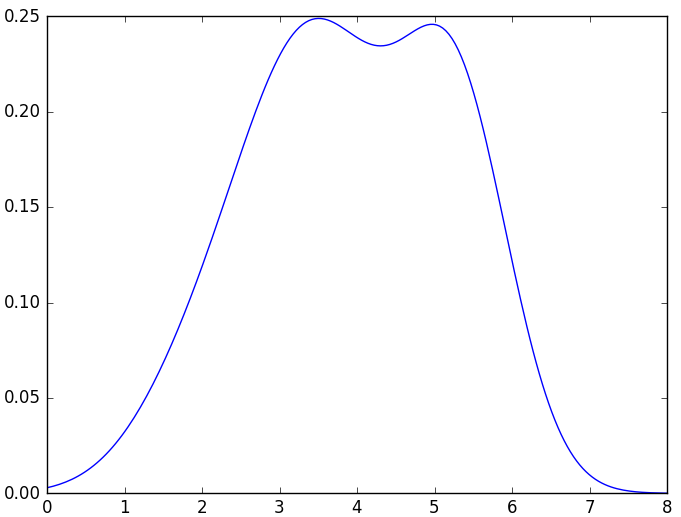}}
		\put(0,20){\rotatebox{90}{\mtiny{Density}}}
		\put(30,0){\mtiny{$y$}}
		\end{picture}
	\end{minipage}
	\begin{minipage}[t]{0.24\linewidth}
		\begin{picture}(50,50)
		\put(4,4){\includegraphics[width=0.92\linewidth]{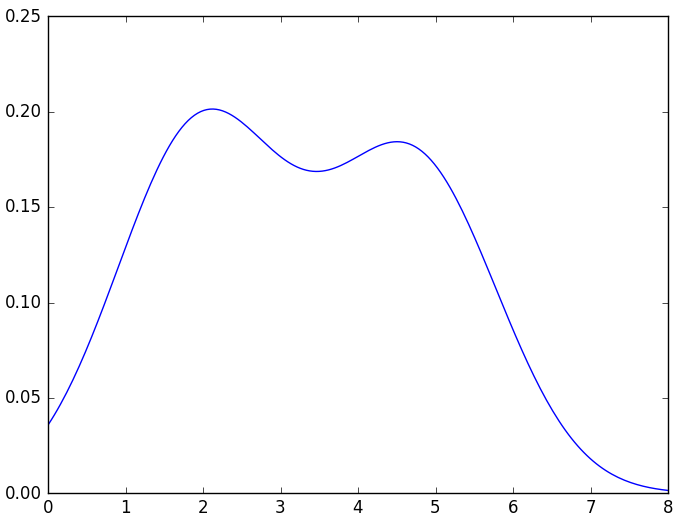}}
		\put(0,20){\rotatebox{90}{\mtiny{Density}}}
		\put(30,0){\mtiny{$y$}}
		\end{picture}
	\end{minipage}
	\begin{minipage}[t]{0.24\linewidth}
		\begin{picture}(50,50)
		\put(4,4){\includegraphics[width=0.92\linewidth]{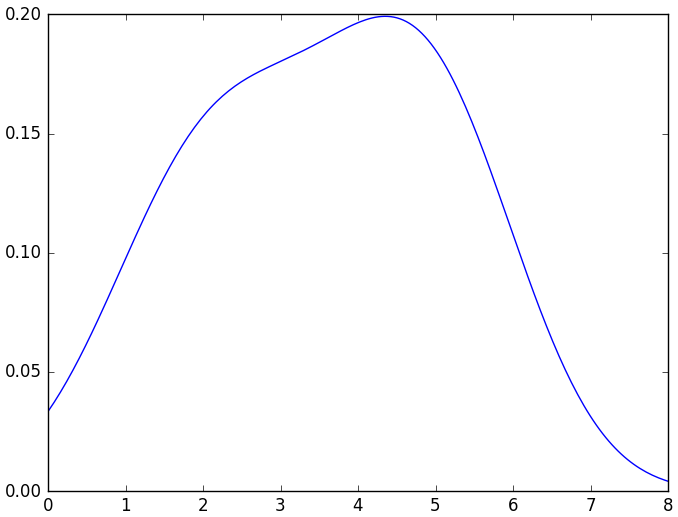}}
		\put(0,20){\rotatebox{90}{\mtiny{Density}}}
		\put(30,0){\mtiny{$y$}}
		\end{picture}
	\end{minipage}
	\begin{minipage}[t]{0.24\linewidth}
		\begin{picture}(50,50)
		\put(4,4){\includegraphics[width=0.92\linewidth]{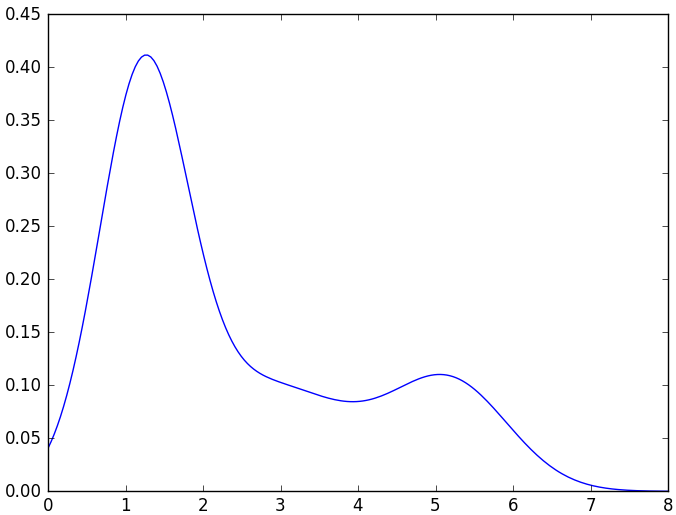}}
		\put(0,20){\rotatebox{90}{\mtiny{Density}}}
		\put(30,0){\mtiny{$y$}}
		\end{picture}
	\end{minipage}
	\begin{minipage}[t]{0.24\linewidth}
		\begin{picture}(50,50)
		\put(4,4){\includegraphics[width=0.92\linewidth]{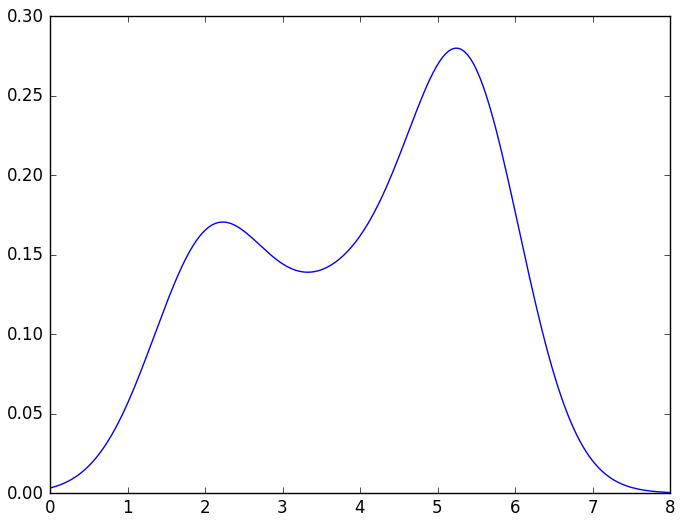}}
		\put(0,20){\rotatebox{90}{\mtiny{Density}}}
		\put(30,0){\mtiny{$y$}}
		\end{picture}
	\end{minipage}
	\begin{minipage}[t]{0.24\linewidth}
		\begin{picture}(50,50)
		\put(4,4){\includegraphics[width=0.92\linewidth]{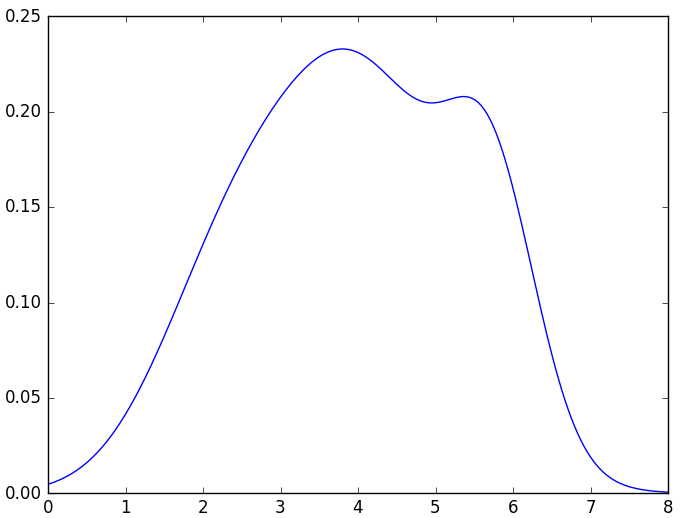}}
		\put(0,20){\rotatebox{90}{\mtiny{Density}}}
		\put(30,0){\mtiny{$y$}}
		\end{picture}
	\end{minipage}
	\caption{Visualization of the learned continuous MoG distributions of representative codes before discretization. It is clear that most of them are multimodal, and therefore cannot be well fitted using a single Gaussian.}
	\label{fig:mog}
\end{figure}

The running time of the competing DNN-based  entropy models is tested on an NVIDIA TITAN Xp machine, whose results on the Kodak dataset are listed in Table~\ref{tab:lossless_time}. For encoding, CCN$_\mathrm{light}$ enjoys the fastest speed, followed by Multiscale PixelCNN and SIN (the best performing variant).
Despite similar encoding time, they have substantially different decoding complexities. Multiscale PixelCNN has the fastest decoder, followed by SIN. Due to the sequential decoding nature, MCN is the slowest, taking nearly one hour to decode a grayscale image of size $752\times496$. Our CCN achieves a significant improvement upon MCN with the proposed code dividing technique, while maintaining nearly the same bit rates.
Moreover, CCN$_\mathrm{light}$ speeds up CCN more than $30$ times, striking a good balance  between model efficiency and model accuracy.
\cmt{We also report the model complexity of the competing models in terms of the number of parameters and floating point operations (FLOPs) in Table~\ref{tab:lossless_time}. Except for the CCN$_\mathrm{light}$, all the competing models have comparable complexity.}

We conduct thorough ablation experiments to analyze the impact of individual components of CCN to  final compression performance. Fig.~\ref{fig:lossless_cmp2} shows the bit rates of CCNs  with three different numbers of feature blocks  ($N\in\{4,8,16\}$) and two filter sizes ($S\in\{3,5\}$). When the filter size and the network depth are fixed, adding more feature blocks effectively increases model capability and thus boosts the compression performance. Similarly, using a larger filter size with fixed network depth and feature block number increases the partial context, leading to better entropy modeling.  Moreover, we replace the proposed zigzag coding order in CCN with the raster coding order, whose model is denoted by CCN$_\mathrm{raster}$. From Fig.~\ref{fig:lossless_cmp2}, we observe that
the performance of CCN$_\mathrm{raster}$ drops significantly, only comparable to the CCN with four feature blocks. This verifies the advantages of the proposed coding order.

\subsection{Lossy Image Compression}
In lossy image compression, the analysis transform $g_a$, the non-uniform quantizer $g_d$, the CCN-based entropy model, and the synthesis transform $g_s$ are jointly optimized for rate-distortion performance. In the early stages of training, the probability $P({\bm y})$ often changes rapidly, which makes it  difficult to keep track of, and may cause instability in learning the entropy model. We find that this issue can be alleviated by a simple warmup strategy. Specifically, $g_a$ and $g_s$ are trained using the distortion term $\ell_d$ only for the first epoch. We then fix $g_a$ and train the CCN-based entropy model until it reasonably fits the current distribution of the codes. After that, we end-to-end optimize the entire method for the remaining epochs. We use Adam with a learning rate of $10^{-5}$ and gradually lower it by a factor of $10$, until $10^{-7}$. The number of quantization centers is $L=8$, and the number of
Gaussian components in MoG is $C=3$.
{
%Compared with the single zero mean Gaussian distribution~\cite{balle2018variational} or linear piecewise functions~\cite{balle2016end} used in the previous lossy image compression methods, the adopted MoG distributions are more accurate for fitting the real distribution of the codes.
Fig.~\ref{fig:mog} shows the learned continuous distributions of representative codes, which are typically multimodal, and therefore cannot be well fitted by unimodal distributions (\textit{e.g.}, Gaussian).}
The optimization is performed separately for each  $\lambda$ and for each distortion measure. We optimize twelve models for six bit rates and two distortion metrics (MSE and MS-SSIM). MSE is converted to peak signal-to-noise ratio (PSNR) for quantitative analysis.

\begin{figure}[!tbp]
	\centering
%	\begin{minipage}[t]{0.45\linewidth}
%		\includegraphics[width=1.0\linewidth]{cmp_psnr}
%		\centering{\scriptsize{(a)}}
%	\end{minipage}
%	\begin{minipage}[t]{0.46\linewidth}
%		\includegraphics[width=1.0\linewidth]{cmp_ms_ssim}
%		\centering{\scriptsize{(b)}}
%	\end{minipage}
	\includegraphics[width=0.85\linewidth]{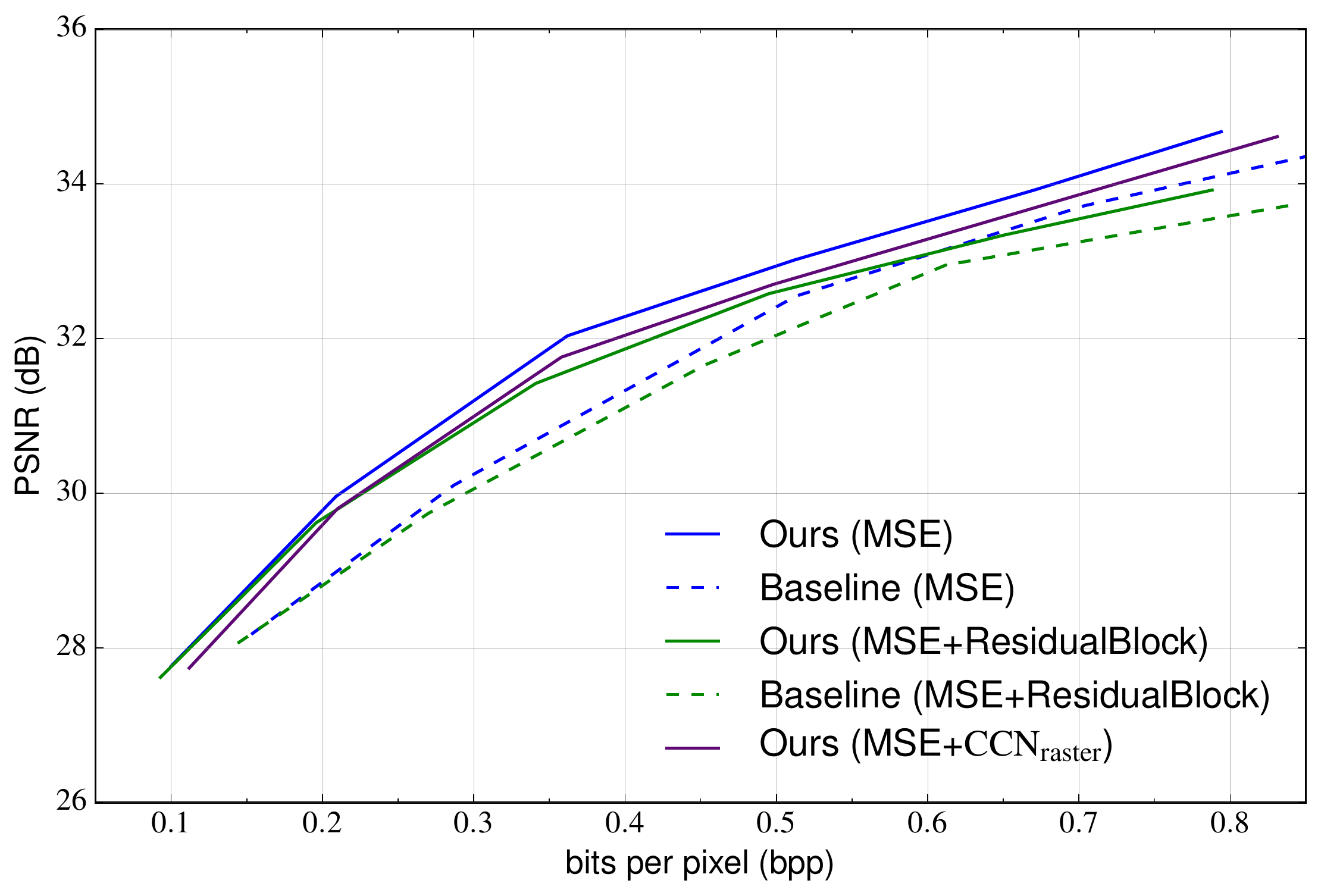}
	\caption{Rate-distortion curves of several variants of the proposed lossy image compression method on the Kodak dataset. }\label{fig:cmp}
\end{figure}

\cmt{
We also evaluate the effect of several components, \ie, the transforms, entropy model, and joint optimization, of the proposed lossy image compression method. All the variants of our method are optimized for MSE distortion and tested on the Kodak dataset. The compression performance of these methods is shown in Fig.~\ref{fig:cmp}.
In terms of the transforms, we replace each dense block of our method with three residual blocks. To keep the same network depth, the first residual block consists of three convolution layers, while the last two residual blocks have two convolution layers. Our method performs better than this variant, especially at higher bit rates.
As for the entropy model, we replace the proposed CCN  with CCN$_\mathrm{raster}$. Similar to the results in lossless image compression, our method with the zigzag coding order outperforms this variant with the raster coding order.
Moreover, we investigate the effect of joint optimization of the transforms and the entropy model. A baseline method is introduced, which first optimizes the transforms for MSE ($\lambda=\infty$), and then trains the CCN-based entropy model by fixing the learned transforms. The baseline method performs worse than the jointly optimized one by 1 dB. Similar trends can be observed for the variants with residual blocks.
}

\begin{figure}[!tbp]
	\centering
	\begin{minipage}[t]{0.45\linewidth}
		\includegraphics[width=1.0\linewidth]{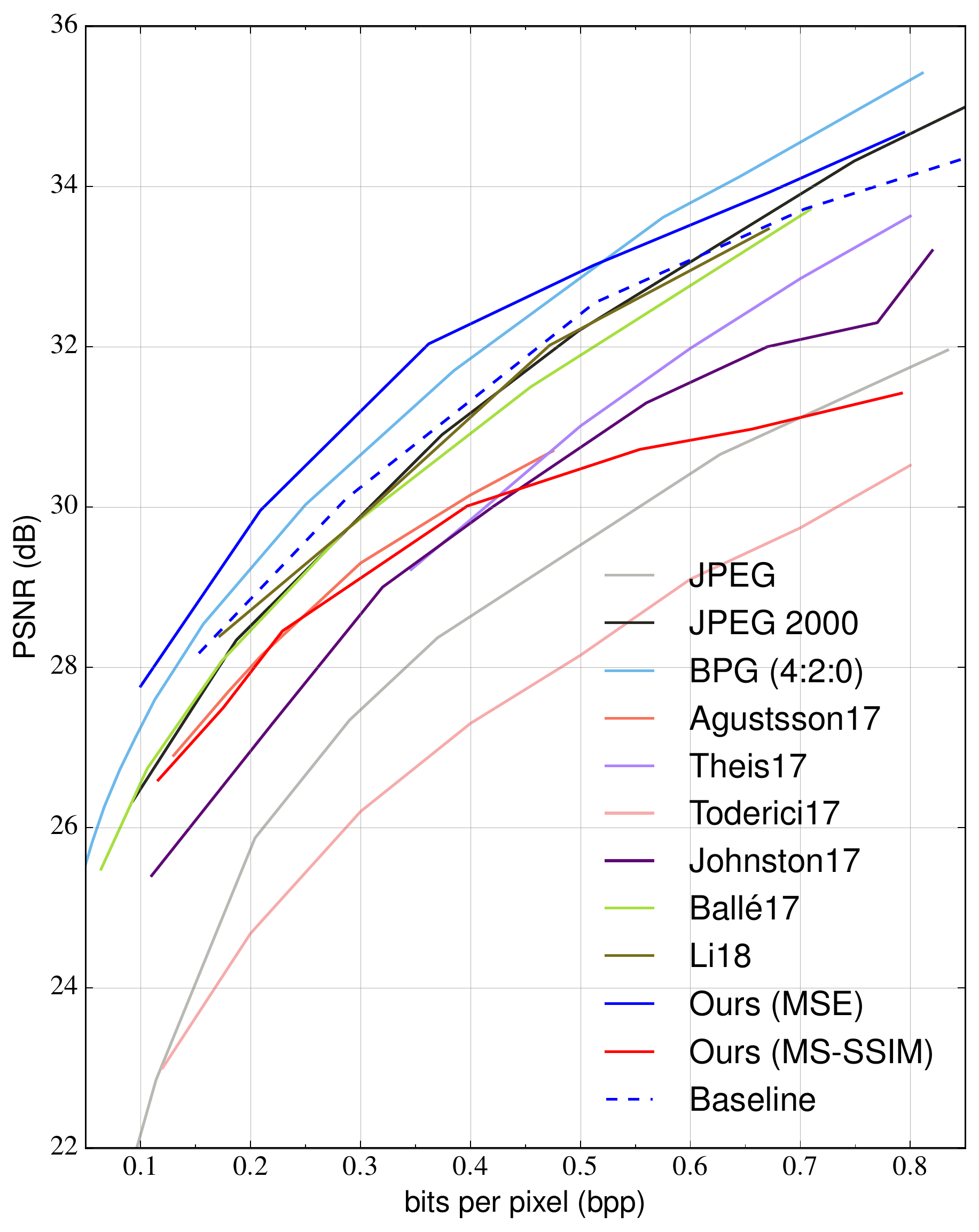}
		\centering{\scriptsize{(a)}}
	\end{minipage}
	\begin{minipage}[t]{0.46\linewidth}
		\includegraphics[width=1.0\linewidth]{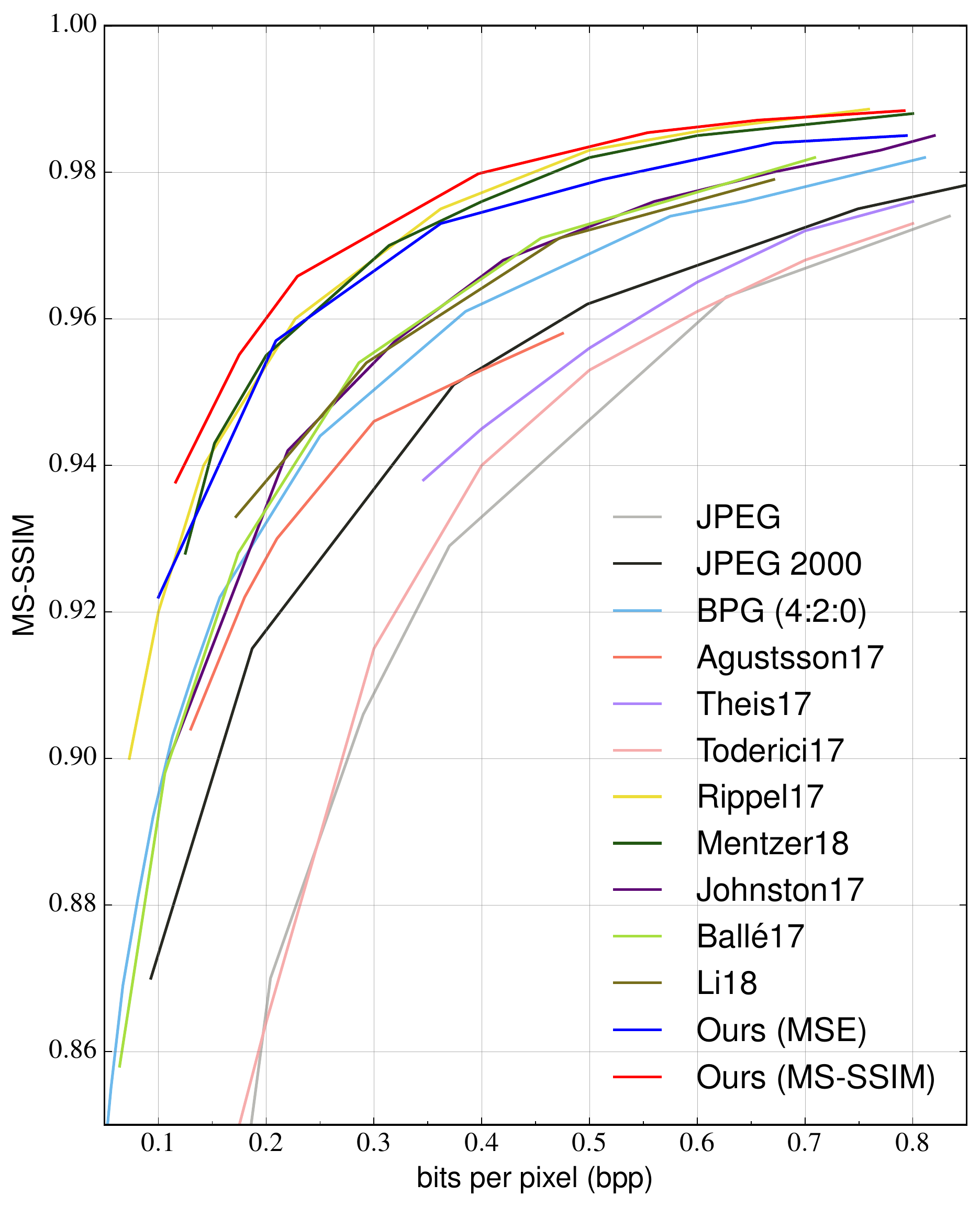}
		\centering{\scriptsize{(b)}}
	\end{minipage}
	\caption{Rate-distortion curves of different compression methods on the Kodak dataset. (a) PSNR. (b) MS-SSIM.}
	\label{fig:kodak}
\end{figure}

\begin{figure}[!tbp]
	\centering
	\begin{minipage}[t]{0.45\linewidth}
		\includegraphics[width=1.0\linewidth]{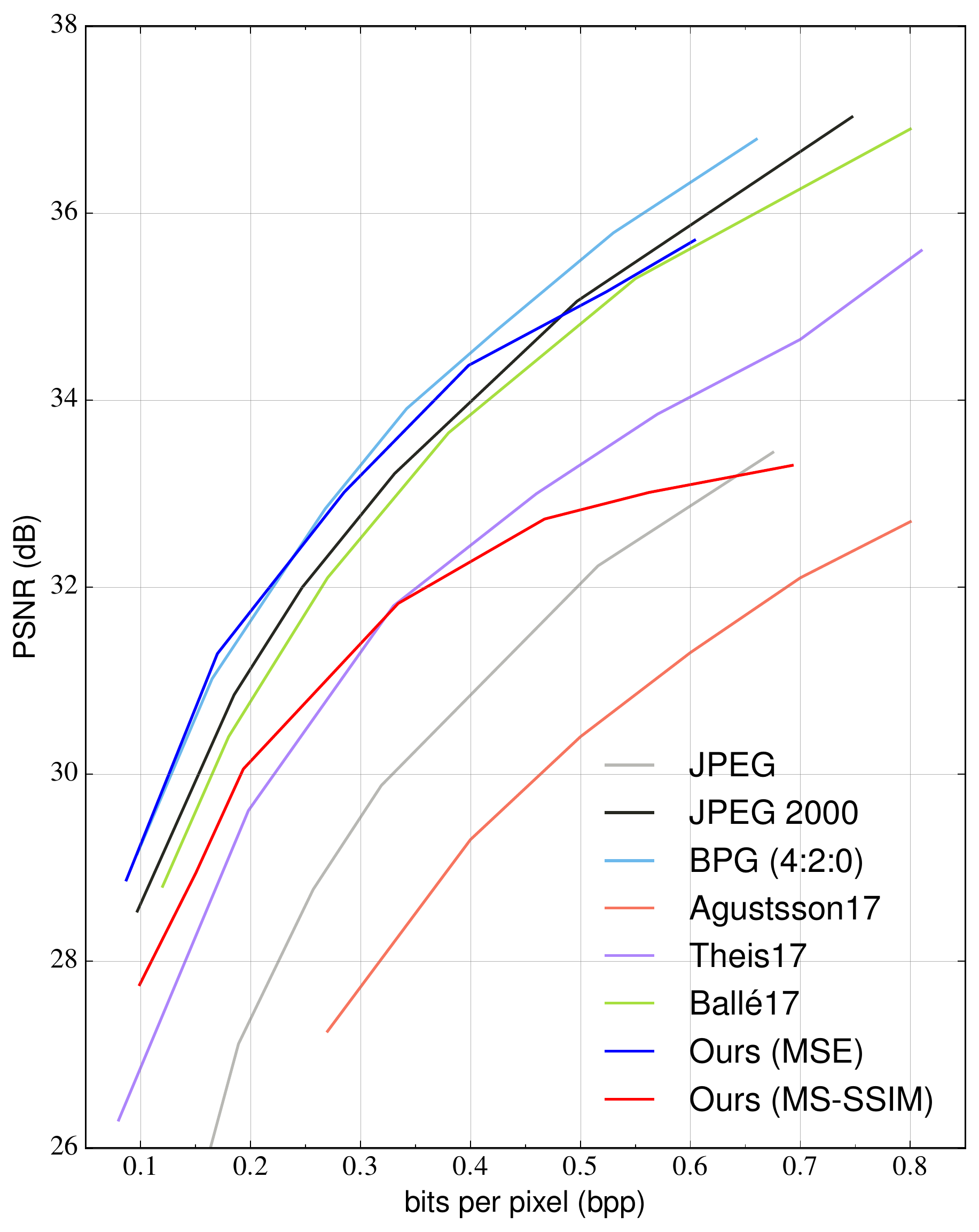}
		\centering{\scriptsize{(a)}}
	\end{minipage}
	\begin{minipage}[t]{0.46\linewidth}
		\includegraphics[width=1.0\linewidth]{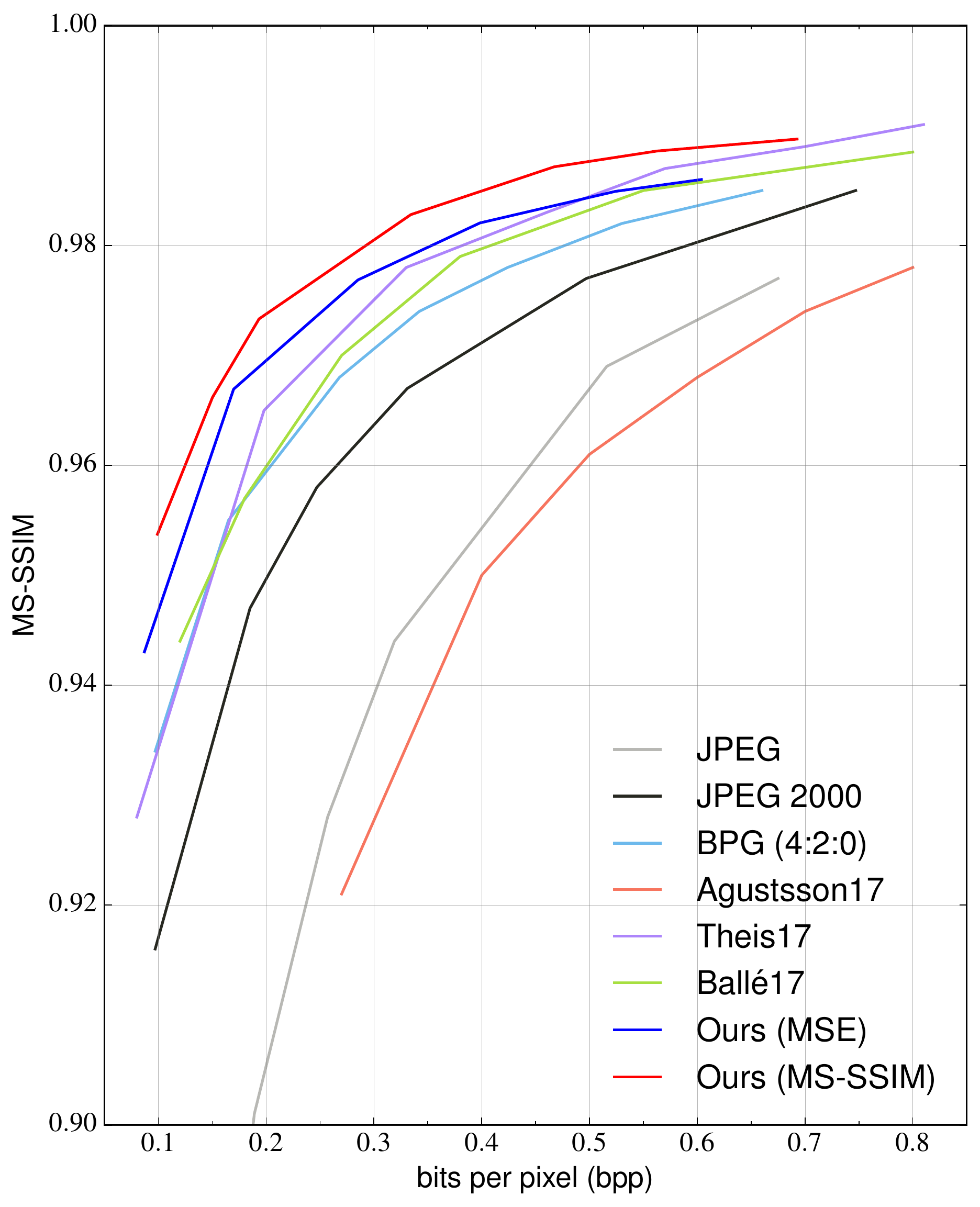}
		\centering{\scriptsize{(b)}}
	\end{minipage}
	\caption{Rate-distortion curves of different compression methods on the Tecnick dataset. (a) PSNR. (b) MS-SSIM.}
	\label{fig:tecnick}
\end{figure}

\begin{figure*}[!tbp]
	\centering
	%\hfill\vline\hfill
	%\begin{minipage}{1.0\textwidth}
	%\begin{minipage}{0.2\textwidth}\center{Original}\end{minipage}
	%\begin{minipage}{0.195\textwidth}\center{Ball{\'e}17~\cite{balle2016end}}\end{minipage}
	%\begin{minipage}{0.195\textwidth}\center{Li18~\cite{li2017learning}}\end{minipage}
	%\begin{minipage}{0.195\textwidth}\center{BPG}\end{minipage}
	%\begin{minipage}{0.195\textwidth}\center{Ours(MS-SSIM)}\end{minipage}
	%\end{minipage}
	%\hfill
	%\hfill\vline\hfill
	\begin{minipage}{1.0\linewidth}
	\includegraphics[width=1.0\linewidth]{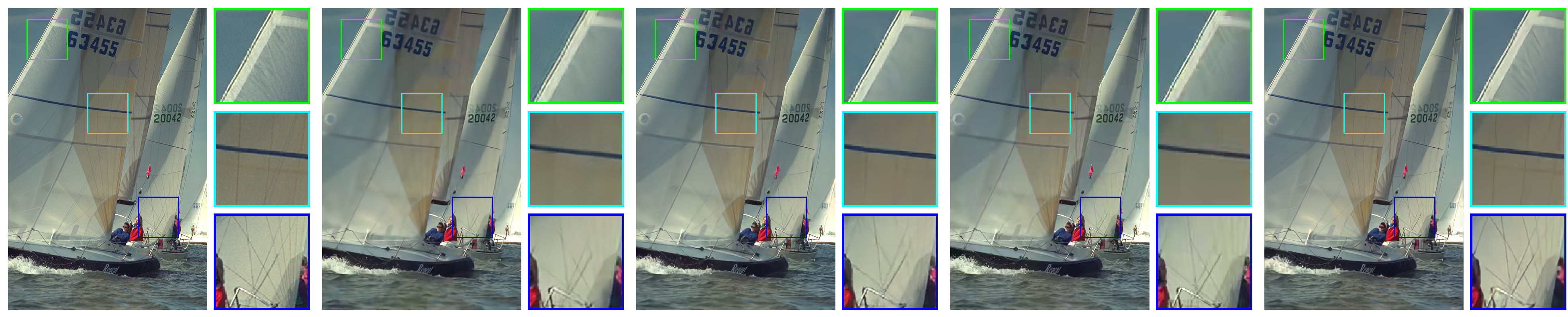}
	\end{minipage}
	%\hfill
	\hfill\vline\hfill
	\begin{minipage}{1.0\textwidth}
	\begin{minipage}{0.2\textwidth}\center{\scriptsize{(a)}}\end{minipage}
	\begin{minipage}{0.195\textwidth}\center{\scriptsize{(b)}}\end{minipage}
	\begin{minipage}{0.195\textwidth}\center{\scriptsize{(c)}}\end{minipage}
	\begin{minipage}{0.195\textwidth}\center{\scriptsize{(d)}}\end{minipage}
	\begin{minipage}{0.195\textwidth}\center{\scriptsize{(e)}}\end{minipage}
	\end{minipage}
	%\hfill
	\hfill\vline\hfill
	\vspace{-.3cm}
	\begin{minipage}{1.0\linewidth}
	\includegraphics[width=1.0\linewidth]{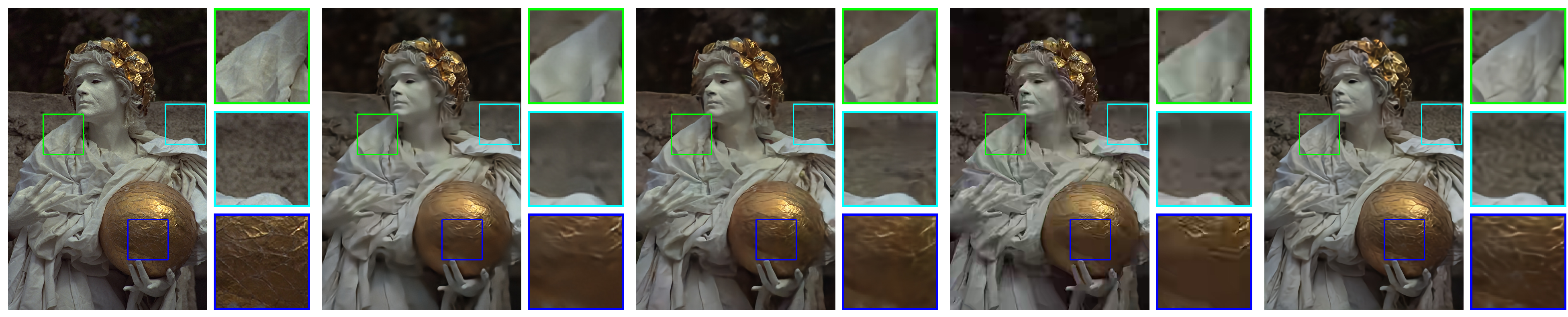}
	\end{minipage}
	%\hfill
	\hfill\vline\hfill
	\begin{minipage}{1.0\textwidth}
	\begin{minipage}{0.2\textwidth}\center{\scriptsize{(f)}}\end{minipage}
	\begin{minipage}{0.195\textwidth}\center{\scriptsize{(g)}}\end{minipage}
	\begin{minipage}{0.195\textwidth}\center{\scriptsize{(h)}}\end{minipage}
	\begin{minipage}{0.195\textwidth}\center{\scriptsize{(i)}}\end{minipage}
	\begin{minipage}{0.195\textwidth}\center{\scriptsize{(j)}}\end{minipage}
	\end{minipage}

\caption{Compressed images by different compression methods on the Kodak dataset. The quantitative measures are in the format of ``bpp / PSNR / MS-SSIM''. (a) Uncompressed ``Sailboat''  image. (b) Ball{\'e}17~\cite{balle2016end}. 0.209 / 31.81 / 0.962. (c) Li18~\cite{li2017learning}. 0.244 / 31.97 / 0.966. (d) BPG. 0.220 / 33.19 / 0.963. (e) Ours optimized for MS-SSIM. 0.209 / 31.01 / 0.978. (f) Uncompressed ``Statue'' image. (g) Ball{\'e}17. 0.143 / 29.48 / 0.942. (h) Li18. 0.115 / 29.35 / 0.938. (i) BPG. 0.119 / 29.77 / 0.935. (j) Ours optimized for MS-SSIM. 0.116 / 28.05 / 0.954. }\label{fig:visual_kodak}
\end{figure*}

 We compare our methods with  existing image coding standards and  recent DNN-based compression models. These include JPEG~\cite{wallace1992jpeg}, JPEG2000~\cite{skodras2001jpeg}, BPG~\cite{bellard2016bpg}, Agustsson17~\cite{agustsson2017soft}, Theis17~\cite{theis2017lossy}, Toderici17~\cite{toderici2016full}, Rippel17~\cite{rippel2017real}, Mentzer18~\cite{mentzer2018conditional1}, Johnston17~\cite{johnston2017improved}, Ball\'{e}17~\cite{balle2016end}, and Li18~\cite{li2017learning}. Both  JPEG (with 4:2:0 chroma subsampling) and JPEG2000 are based on the optimized implementations in MATLAB2017. For BPG, we adopt the latest version from its official website with the default setting. When it comes to DNN-based models for lossy image compression,  the implementations are generally not available. Therefore, we  copy the results from their respective papers.

\begin{figure*}
	\centering
	%\begin{minipage}{1.0\textwidth}
	%\begin{minipage}{0.2\textwidth}\center{Original}\end{minipage}
	%\begin{minipage}{0.195\textwidth}\center{JPEG2K}\end{minipage}
	%\begin{minipage}{0.195\textwidth}\center{Li18~\cite{li2017learning}}\end{minipage}
	%\begin{minipage}{0.195\textwidth}\center{BPG}\end{minipage}
	%\begin{minipage}{0.195\textwidth}\center{Ours(MS-SSIM)}\end{minipage}
	%\end{minipage}
	\hfill
	% \hfill\vline\hfill
	% \begin{minipage}{1.0\linewidth}
	% \includegraphics[width=1.0\linewidth]{tec_1}
	% \end{minipage}
	% \begin{minipage}{1.0\textwidth}
	% \begin{minipage}{0.2\textwidth}\center{\footnotesize {bpp / PSNR(dB) / MS-SSIM}}\end{minipage}
	% \begin{minipage}{0.195\textwidth}\center{\footnotesize{0.147 / 29.773 / 0.956}}\end{minipage}
	% \begin{minipage}{0.195\textwidth}\center{\footnotesize{0.140 / 27.661 / 0.954}}\end{minipage}
	% \begin{minipage}{0.195\textwidth}\center{\footnotesize{0.153 / 31.831 / 0.970}}\end{minipage}
	% \begin{minipage}{0.195\textwidth}\center{\footnotesize{0.147 / 28.698 / 0.978}}\end{minipage}
	% \end{minipage}
	%\hfill\vline\hfill
	\begin{minipage}{1.0\linewidth}
	\includegraphics[width=1.0\linewidth]{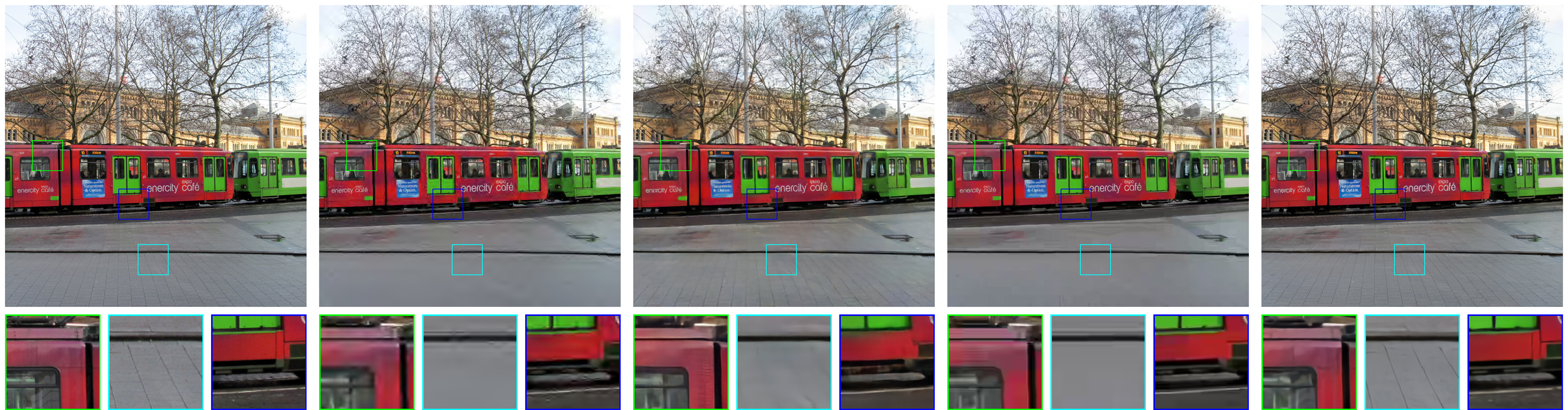}
	\end{minipage}
	%\hfill
	\hfill\vline\hfill
	\begin{minipage}{1.0\textwidth}
	\begin{minipage}{0.2\textwidth}\center{\scriptsize{(a)}}\end{minipage}
	\begin{minipage}{0.195\textwidth}\center{\scriptsize{(b)}}\end{minipage}
	\begin{minipage}{0.195\textwidth}\center{\scriptsize{(c)}}\end{minipage}
	\begin{minipage}{0.195\textwidth}\center{\scriptsize{(d)}}\end{minipage}
	\begin{minipage}{0.195\textwidth}\center{\scriptsize{(e)}}\end{minipage}
	\end{minipage}
	%\hfill
	\hfill\vline\hfill
	\vspace{-.3cm}
	\begin{minipage}{1.0\linewidth}
	\includegraphics[width=1.0\linewidth]{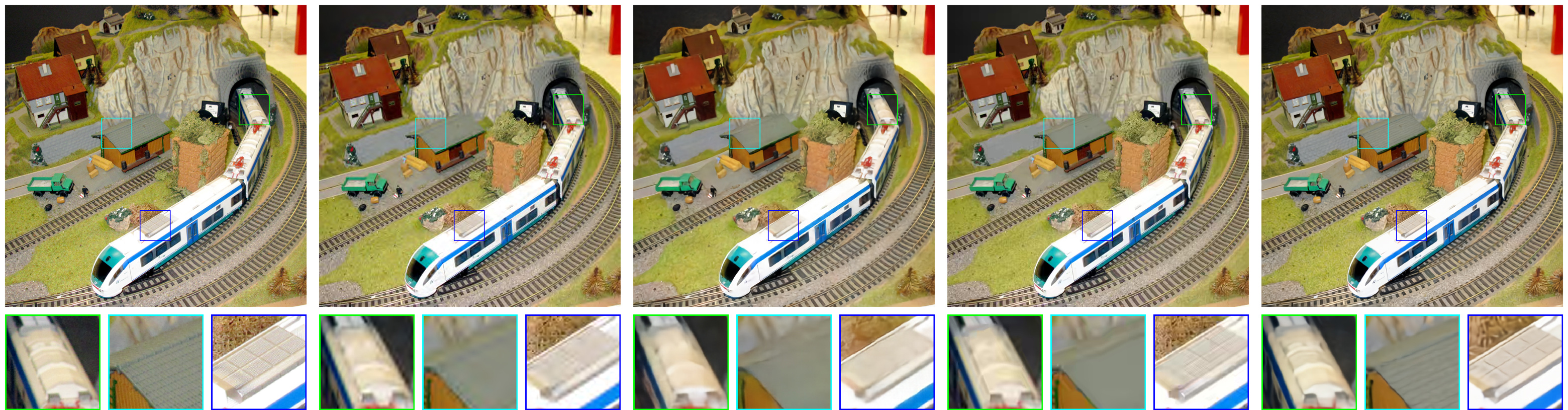}
	\end{minipage}
	%\hfill
	\hfill\vline\hfill
	\begin{minipage}{1.0\textwidth}
	\begin{minipage}{0.2\textwidth}\center{\scriptsize{(f)}}\end{minipage}
	\begin{minipage}{0.195\textwidth}\center{\scriptsize{(g)}}\end{minipage}
	\begin{minipage}{0.195\textwidth}\center{\scriptsize{(h)}}\end{minipage}
	\begin{minipage}{0.195\textwidth}\center{\scriptsize{(i)}}\end{minipage}
	\begin{minipage}{0.195\textwidth}\center{\scriptsize{(j)}}\end{minipage}
	\end{minipage}
\caption{Compressed images by different compression methods on the Tecnick dataset. The quantitative measures are in the format of ``bpp / PSNR / MS-SSIM''. (a) Uncompressed ``Bus'' image. (b) JPEG2K. 0.199 / 24.41 / 0.914. (c) Li18~\cite{li2017learning}. 0.224 / 23.41 / 0.908. (d) BPG. 0.208 / 25.36 / 0.928. (e) Ours (MS-SSIM). 0.198 / 23.71 / 0.951. (f) Uncompressed ``Toy train'' image.  (g) JPEG2K. 0.201 / 28.29 / 0.917. (h) Li18. 0.189 / 26.83 / 0.899. (i) BPG. 0.210 / 29.25 / 0.933. (j) Ours (MS-SSIM). 0.198 / 28.08 / 0.949.}\label{fig:visual_tec}
\end{figure*}

Fig.~\ref{fig:kodak} shows the rate-distortion curves on the Kodak dataset. We find that our method optimized for MSE outperforms all competing methods at low bit rates ($<0.5$ bpp), except for BPG. When optimized for MS-SSIM, our method performs on par with Rippel17 and is much better than the rest. Fig.~\ref{fig:tecnick} shows the rate-distortion curves on the Tecnick dataset, where we observe similar trends for both PSNR and MS-SSIM. An interesting observation is that when we continue increasing the bit rate, PSNR/MS-SSIM starts to  plateau, which may be due to  the limited model capability. Without any constraint on rate ($\lambda = \infty$) and quantization ($L=\infty$), our method optimized for MSE only reaches $38.2$ dB on the Kodak dataset, which can be treated as an empirical upper-bound for our network structure. Preliminary results from~\cite{balle2018variational} indicate that increasing the depth and width of the network leads to  performance improvements at high bit rates.

We visually compare the compressed images by our method against Ball{\'e}17, Li18, JPEG2K, and BPG. Fig.~\ref{fig:visual_kodak} and Fig.~\ref{fig:visual_tec} show  sample compressed results on the Kodak and Tecnick datasets, respectively.  JPEG2K and BPG  exhibit artifacts (such as blocking, ringing, blurring, and aliasing) that are common to all handcrafted transform coding methods, reflecting the underlying linear basis functions. Ball{\'e}17 is effective at suppressing ringing artifacts at the cost of over-smoothing fine structures. Li18 allocates more bits to preserve large-scale strong edges, while tends to eliminate small-scale localized features (\eg, edges, contours, and textures). In contrast, our method generates compressed images with more faithful details and less visible distortions.

We report the running time of our method at six bit rates on the Kodak dataset using the same machine. It takes $0.024$ second to generate ${\bm y}$ and $0.032$ second to reconstruct the image. The entropy coding time is listed in Table~\ref{tab:time_lossy}, where we see that more time is needed to encode and decode images at higher bit rates. This is because,
{
in our experiments, we use a larger $M$ at higher bit rates as a way of preserving more perceptually meaningful information. This corresponds to more convolution operations and longer processing time.
}
With the help of the proposed code dividing technique, our method performs entropy decoding in around one second for images of size $752\times496$.
{
We also note that the entropy coding time for lossy image compression is much faster than that of lossless image compression for two main reasons. First, in lossless compression, the size of the code block to CCN for entropy coding is $8 \times H \times W$. In lossy compression, the input image is first transformed to the code block with the size of $M\times\left(H/8\right)\times\left(W/8\right)$ ($M\le32$). With a smaller code block size, the lossy CCN is expected to be faster than the lossless CCN. Moreover, $N$ and $S$ also highly affect the speed. For the CCN in lossless compression, we set $N=16$ and $S=5$, whereas for the CCN in lossy compression, we set $N=3$ and $S=5$, leading to the faster speed.
}

\begin{table}
	\centering
	\caption{Running time in seconds  of our CCN-based entropy model at six bit rates on the Kodak dataset}
	\begin{tabular}{ccccccc}
		\toprule
		Average bpp&0.100&0.209&0.362&0.512&0.671&0.794\\
		\midrule
		Encoding&0.013&0.025&0.044&0.066&0.085&0.103\\
		Decoding&0.116&0.227&0.457&0.735&1.150&1.232\\
		\bottomrule
	\end{tabular}
	
	\label{tab:time_lossy}
\end{table}

\section{Conclusion and Discussion}
We have introduced CCNs for context-based entropy modeling. Parallel entropy encoding and decoding are achieved with the proposed  coding order and code dividing technique, which can be efficiently implemented using masked convolutions. We tested the CCN-based entropy model (combined with  arithmetic coding) in both lossless and lossy image compression. For the lossless case, our method achieves the best compression performance, which we believe arises from the more accurate estimation of the Bernoulli distributions of the binary codes. For the lossy case, our method offers improvements both visually and in terms of rate-distortion performance over image compression standards and recent DNN-based models.

The application scope of the proposed CCN is far beyond building the entropy model in image compression. As a general probability model, CCN appears promising for a number of image processing applications. For example,  we may use CCN to learn a probability model $P(\bm x)$ for natural images, and use it as a prior in Bayesian inference to solve various vision tasks such as image restoration~\cite{zhang2017beyond,portilla2003image}, image quality assessment~\cite{Wang2006,ma2018}, and image generation\cite{oord2016pixel,chen2018neural}.

%\nocite{*}

\section*{Acknowledgment}
This work was supported in part by Shenzhen Research Institute of Big Data and Shenzhen Institute of Artificial Intelligence and Robotics for Society. The authors would like to thank the NVIDIA Corporation for donating a TITAN Xp GPU used in this research.

%\begin{thebibliography}{1}

\bibliographystyle{IEEEtran}
\bibliography{IEEEabrv,./egbib}
%
%%\end{thebibliography}
\begin{IEEEbiography}
[{\includegraphics[width=1in,height=1.25in,clip,keepaspectratio]{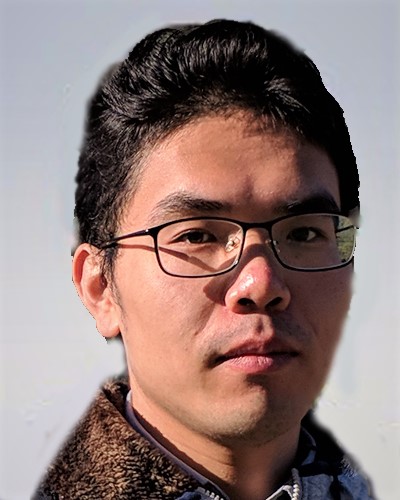}}]{Mu Li} received the B.C.S. degree in Computer Science and Technology from Harbin Institute of Technology, Harbin, China, in 2015. He is the owner of the Hong Kong PhD Fellowship and is currently working toward the Ph.D. degree in Department of Computing from the Hong Kong Polytechnic University, Hong Kong, China. His research interests include deep learning and image processing.
\end{IEEEbiography}

\begin{IEEEbiography}[{\includegraphics[width=1in,height=1.25in,clip,keepaspectratio]{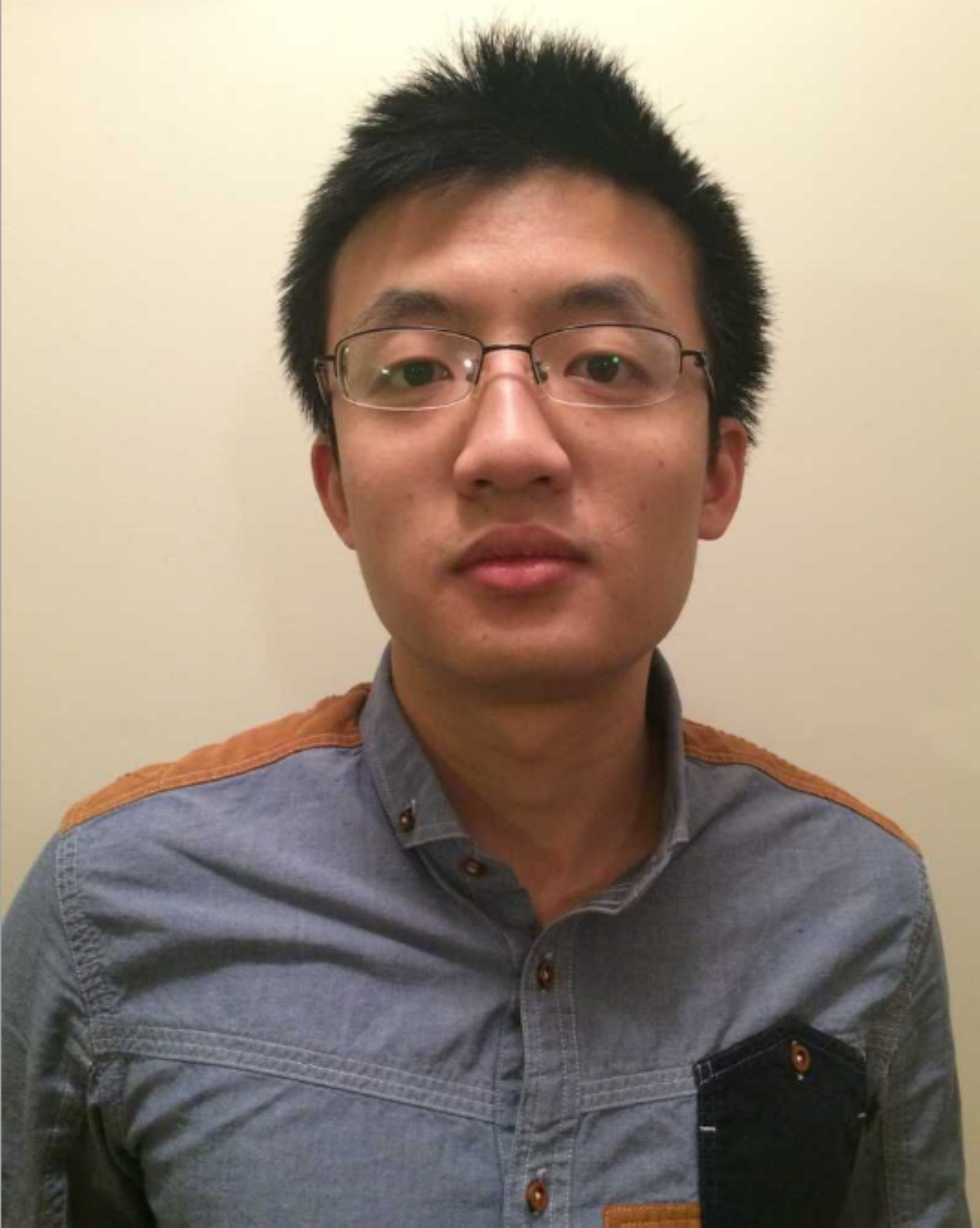}}]{Kede Ma}
(S'13-M'18) received the B.E. degree from the University of Science and Technology of China, Hefei, China, in 2012, and the M.S. and Ph.D.
degrees in electrical and computer engineering from the University of Waterloo, Waterloo, ON, Canada, in 2014 and 2017, respectively. He was a Research Associate with the Howard Hughes Medical Institute and New York University, New York, NY, USA, in 2018. He is currently an Assistant Professor with the Department of Computer Science, City University of Hong Kong. His research interests include perceptual image processing, computational vision, and computational photography.
\end{IEEEbiography}
\begin{IEEEbiography}
[{\includegraphics[width=1in,height=1.25in,clip,keepaspectratio]{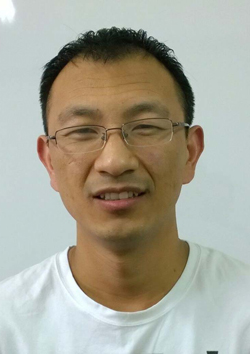}}]{Wangmeng Zuo}(M'09-SM'14) received the Ph.D. degree in computer application technology from the Harbin Institute of Technology, Harbin, China, in 2007. From 2004 to 2006, he was a Research Assistant with the Department of Computing, The Hong Kong Polytechnic University, Hong Kong. From 2009 to 2010, he was a Visiting Professor with Microsoft Research Asia. He is currently a Professor with the School of Computer Science and Technology, Harbin Institute of Technology. He has published over 60 papers in top-tier academic journals and conferences. His current research interests include image enhancement and restoration, weakly supervised learning, visual tracking, and image classification. He has served as a Tutorial Organizer in ECCV 2016, an Associate Editor of the IET Biometrics, and the Guest Editor of Neurocomputing, Pattern Recognition, the IEEE TRANSACTIONS ON CIRCUITS AND SYSTEMS FOR VIDEO TECHNOLOGY, and the IEEE TRANSACTIONS ON NEURAL NETWORKS AND LEARNING SYSTEMS.
\end{IEEEbiography}
\begin{IEEEbiography}
[{\includegraphics[width=1in,height=1.25in,clip,keepaspectratio]{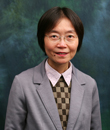}}]{Jane You} received the B.Eng. degree in electronics engineering from Xi'an Jiaotong University, Xi'an, China, in 1986, and the Ph.D. degree in computer science from La Trobe University, Melbourne, VIC, Australia, in 1992. She was a Lecturer with the University of South Australia, Adelaide SA, Australia, and a Senior Lecturer with Griffith University, Nathan, QLD, Australia, from 1993 to 2002. She is currently a Full Professor with The Hong Kong Polytechnic University, Hong Kong. Her current research interests include image processing, pattern recognition, medical imaging, biometrics computing, multimedia systems, and data mining.
\end{IEEEbiography}
\begin{IEEEbiography}
[{\includegraphics[width=1in,height=1.25in,clip,keepaspectratio]{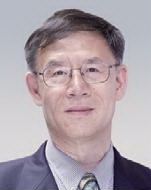}}]{David Zhang} received the Graduation degree in computer science from Peking University, Beijing, China, the M.Sc. degree in computer science in 1982, and the Ph.D. degree in 1985 from the Harbin Institute of Technology (HIT), Harbin, China. In 1994, he received the second Ph.D. degree in electrical and computer engineering from the University of Waterloo, Waterloo, Ontario, Canada. From 1986 to 1988, he was a Postdoctoral Fellow with Tsinghua University and then an Associate Professor at the Academia Sinica, Beijing. He is currently a Chair Professor at the Hong Kong Polytechnic University and the Chinese University of Hong Kong (Shenzhen). He also serves as a Visiting Chair Professor at Tsinghua University, Beijing, and an Adjunct Professor at Peking University, Shanghai Jiao Tong University, Shanghai, China, HIT, and the University of Waterloo. His research interests are medical biometrics and pattern recognition.
\end{IEEEbiography}

\end{document}